\documentclass[a4paper,fleqn,usenatbib]{mnras}
\usepackage{newtxtext,newtxmath}
\usepackage[T1]{fontenc}
\usepackage{ae,aecompl}
\usepackage[dvipdfmx]{graphicx}
\usepackage{amsmath}
\usepackage{amssymb}
\usepackage{lscape}
\usepackage{graphicx}

\title[Fragmentation of discs around SMSs]{Gravitational stability and fragmentation condition for discs around accreting supermassive stars}

\author[R. Matsukoba et al.]{
Ryoki Matsukoba,$^{1}$\thanks{E-mail: r.matsukoba@astr.tohoku.ac.jp}
Sanemichi Z. Takahashi,$^{2,3}$
Kazuyuki Sugimura$^{1}$ and
\newauthor
~Kazuyuki Omukai$^{1}$
\\
$^{1}$Astronomical Institute, Tohoku University, Aoba, Sendai, Miyagi 980-8578, Japan\\
$^{2}$Department of Applied Physics, Kogakuin University, Nakano, Hachioji, Tokyo 192-0015, Japan\\
$^{3}$National Astronomical Observatory of Japan, Osawa, Mitaka, Tokyo 181-8588, Japan
}

\date{Accepted XXX. Received YYY; in original form ZZZ}

\pubyear{2018}

\begin{document}
\label{firstpage}
\pagerange{\pageref{firstpage}--\pageref{lastpage}}
\maketitle

\begin{abstract}
Supermassive stars (SMSs) with mass $\sim10^{5}~\mathrm{M}_{\odot}$ are
promising candidates for the origin of supermassive black holes observed
at redshift $\gtrsim6$.  They are supposed to form as a result of rapid
accretion of primordial gas, although it can be obstructed by the time
variation caused by circum-stellar disc fragmentation due to
gravitational instability.
To assess the occurrence of fragmentation, we study the structure of
marginally gravitationally unstable accretion discs, by using a steady
one-dimensional thin disc model with detailed treatment of chemical and thermal
processes. Motivated by two SMS formation scenarios, i.e., those with strong ultraviolet
radiation background or with large velocity difference between the baryon
and the dark matter, we consider two types of flows, i.e., atomic and
molecular flows, respectively, for a wide range of 
the central stellar mass $10-10^5~\mathrm{M}_{\odot}$ and the accretion rate 
$10^{-3}-1~\mathrm{M}_{\odot}~\mathrm{yr}^{-1}$. 
In the case of a mostly atomic gas flowing to the disc outer boundary, 
the fragmentation condition is expressed as 
the accretion rate being higher than 
the critical value of $10^{-1}~\mathrm{M}_{\odot}~\mathrm{yr}^{-1}$ regardless of the central stellar mass.
On the other hand, in the case of molecular flows, 
there is a critical disc radius outside of which the disc becomes unstable. 
Those conditions appears to be marginally satisfied according to 
numerical simulations, suggesting that disc fragmentation can 
be common during SMS formation.

\end{abstract}



\section{Introduction}
\label{Sec:introduction}

Discoveries of dozens of quasars at high redshift $z \gtrsim 6.5$ demonstrate the very existence of supermassive black holes (SMBHs) 
as massive as $\sim 10^{9}~\mathrm{M}_{\odot}$ in less than a billion years after the Big Bang 
(e.g., \citealt{Venemans:2013}; \citealt{Matsuoka:2018}; see also \citealt{Gallerani:2017} for a review), 
including the most distant one with $8 \times 10^{8}~\mathrm{M}_{\odot}$ at redshift $z = 7.5$ \citep{Banados:2018}.
Although the origin of those BHs is still a mystery, short available time for their growth seems to favor 
massive seed BHs (see, e.g., \citealt{Volonteri:2012}; \citealt{Haiman:2013} for a review).

Remnant BHs of Population III stars with $\sim10^{2}~\mathrm{M}_{\odot}$ are one of the candidates for such seeds
\citep[e.g.,][]{Madau:2001}. They can grow SMBHs in the available time 
if either with continuous accretion at the Eddington limit or with short
episode of super-Eddington growth. In reality, however, the BH's growth
can easily be hindered by its own radiative feedback 
\citep{Milosavljevic:2009,Park:2011,Sugimura:2018,Orofino:2018}. 
An alternative and attractive pathway for seed BH formation is via the
so-called direct collapse (e.g., \citealt{Bromm:2003}),
where a supermassive star (SMS) with $\sim 10^{5}~\mathrm{M}_{\odot}$ collapses
into a BH with a similar mass by general relativistic instability \citep{Umeda:2016,Woods:2017,Haemmerle:2018}.
Here, SMSs are supposed to form from a primordial gas in some peculiar sites.
Unlike in ordinary first star formation, which is driven by
$\mathrm{H}_{2}$ cooling (see, e.g., \citealt{Bromm:2004};
\citealt{Glover:2013} for a review), $\mathrm{H}_{2}$ is dissociated by
strong external far-ultraviolet (UV) radiation from nearby galaxies and the
contraction of clouds is solely caused by $\mathrm{H}$ atomic cooling in
the most intensively studied channel for SMS formation 
\citep{Omukai:2001,Omukai:2008,Shang:2010,Regan:2014,Sugimura:2014}.
Such clouds contract almost isothermally at $\sim 10^{4}$ K
without experiencing vigorous fragmentation.  The protostar formed at
the centre subsequently accretes the gas at a high rate of $10^{-1}~\mathrm{M}_{\odot}~\mathrm{yr}^{-1}$ due to high temperature
\citep{Latif:2013,Inayoshi:2014-11,Becerra:2015,Chon:2018}. 
Such rapidly accreting protostar inflates greatly in radius with effective
temperature of several 1000 K and grows supermassive with $\gtrsim 10^{5}~\mathrm{M}_{\odot}$ avoiding ionization radiation feedback on the accretion
flow \citep{Hosokawa:2012,Hosokawa:2013}, before collapsing by general
relativistic instability.

In addition to the above scenario for SMS formation, \cite{Hirano:2017}  recently proposed another channel 
where the suppression of $\mathrm{H}_{2}$ cooling is not always required.
They considered a pristine cloud in a halo with 3$\sigma$ higher than the average velocity difference 
between the dark matter and the baryon generated before the recombination epoch at $z \gtrsim 1100$ 
\citep{Tseliakhovich:2010, Fialkov:2012, Schauer:2017}.
As a result of enhanced effective Jeans mass by such streaming motions, 
the formed protostar in the halo accretes a gas
with a high rate of $\sim 10^{-1}~\mathrm{M}_{\odot}~\mathrm{yr}^{-1}$.  
The star eventually grows to $4 \times 10^{4}~\mathrm{M}_{\odot}$ in their calculation.

Protostellar ionization feedback is suppressed as long as the accretion rate higher than 
$10^{-1}~\mathrm{M}_{\odot}~\mathrm{yr}^{-1}$ is maintained \citep{Hosokawa:2012, Hosokawa:2013}.
Once the accretion rate falls below this value for sometime, however, the radiation feedback may become 
significant and eventually terminate the stellar growth. 
Adopting a simple sporadic accretion history with repeating burst and quiescent phases, \cite{Sakurai:2015} showed 
that the protostar contracts to the main sequence and ionization feedback becomes active 
if the quiescent phase exceeds $\sim 10^{3}$ yrs.
Such time variations would be caused by clump formation in the circum-(proto)stellar disc, 
which is formed due to the finite angular momentum of the cloud.
In fact, numerical simulations of SMS formation \citep{Chon:2018,Hirano:2017} found a sign of disc fragmentation by gravitational instability 
and resulting accretion variation although their spatial resolution 
and chemical/thermal modeling are not good enough to draw definitive conclusions. 

\cite{Inayoshi:2014-12} and \cite{Latif:2015} studied the gravitational
stability of circum-stellar discs in the process of SMS formation by way
of the steady thin disc model \citep{Shakura:1973}, and claimed that the
disc becomes unstable and fragmentation occurs for a
disc accreting at the typical rate of $10^{-1} \ \mathrm{M}_{\odot}~\mathrm{yr}^{-1}$.
In their calculation, however, simplified treatment of thermal modeling is
adopted without solving detailed chemical reactions: \cite{Latif:2015}
assumed that the accretion flow is composed of $\mathrm{H}_{2}$ at the
outer boundary and becomes atomic $\mathrm{H}$ once the temperature
exceeds 4000 K, while \cite{Inayoshi:2014-12} assumed that the gas is
always in the atomic state and the temperature is determined by the
balance between the viscous heating and the $\mathrm{H}^{-}$ free-bound
emission cooling.  Their conclusion needs to be justified by more
detailed modeling.

In this paper, we construct models for one-dimensional steady accretion discs by solving non-equilibrium chemical and thermal evolution
and investigate their gravitational stability. 
While the works by \cite{Inayoshi:2014-12} and \cite{Latif:2015} are limited to a typical accretion rate of 
$10^{-1}~\mathrm{M}_{\odot}~\mathrm{yr}^{-1}$, here we also study the cases with different values of 
the accretion rate and central stellar mass. 

The paper is organized as follows.
We describe our disc model and stability criterion in Section~\ref{Sec:model}.
We then present the obtained disc structures and the result for the stability analysis in Section~\ref{Sec:result}.
Discussion and conclusion are given in Section~\ref{Sec:discussion_conclusion}.
Appendices are for the details of radiative and chemical processes.

\section{Model}
\label{Sec:model}

\subsection{Basic Framework}
\label{Sec:basic_framework}
We construct the one-dimensional axisymmetric and steady model for 
an accretion disc feeding the central star 
under a given central stellar mass and accretion rate.
The disc is assumed to rotate at the Keplerian angular velocity
\begin{align}
\Omega = \sqrt{\frac{G M_{\ast}}{r^{3}}} \  ,
\label{eq:angular_velocity}
\end{align}
where $r$, $G$ and $M_{\ast}$ are the radius, the gravitational constant and the central stellar mass, respectively.
We assume the disc is marginally gravitationally unstable and the angular momentum is transported by the spiral arms.
Thermal and chemical evolution in the flow is calculated as in Sections~\ref{sec:thermal} and \ref{sec:chemical}.
We describe the angular momentum transport by the $\alpha$-viscosity prescription \citep{Shakura:1973} 
and judge disc fragmentation from the required value of $\alpha$-parameter (see Section~\ref{Sec:fragmentation_condition}).

For a marginally unstable disc, Toomre's $Q$-value \citep{Toomre:1964} is 
\begin{align}
Q = \frac{c_{\mathrm{s}} \Omega}{\mathrm{\pi} G \Sigma} \sim 1 \  ,
\end{align}
where $c_{\mathrm{s}}$ is the sound velocity and $\Sigma$ is the surface density, 
as a result of the angular momentum transfer by the spiral arms \citep{Vorobyov:2007,Tsukamoto:2015}.
Another possible mechanism of the angular momentum transfer is via the turbulent viscosity.
According to the results shown in Section~\ref{Sec:result}, however, 
the actual $\alpha$-value required is always larger than the typical value for the turbulent viscosity 
($\alpha \sim 10^{-2}$ for magnetorotational instability; e.g., \citealt{Bai:2013}).
This means that the turbulent viscosity alone cannot maintain the disc in a gravitationally stable state
and the disc eventually settles in a marginally unstable state where 
the angular momentum is transferred due to the spiral arms.
From the condition $Q=1$, the surface density is obtained as
\begin{align}
\Sigma = \frac{c_{\mathrm{s}} \Omega}{\mathrm{\pi} G} \  .
\label{eq:density}
\end{align}
The sound velocity is
\begin{align}
c_{\mathrm{s}} = \sqrt{\frac{\gamma k_{\mathrm{B}} T}{\mu m_{\mathrm{H}}}} \  ,
\label{eq:sound_velocity}
\end{align}
where $\gamma$ is the specific heat ratio, $k_{\mathrm{B}}$ is the Boltzmann constant, $T$ is the gas temperature, $\mu$ is the mean molecular weight and $m_{\mathrm{H}}$ is the mass of a hydrogen nucleus.

The gas mass density is given by
\begin{align}
\rho = \frac{\Sigma}{2 H} \  ,
\label{eq:mass_density}
\end{align}
where the scale height $H$ characterizing the thickness of the disc is estimated from vertical hydrostatic balance:
\begin{align}
H = \frac{c_{\mathrm{s}}}{\Omega} \  .
\label{eq:scale_height}
\end{align}
The radial velocity is related to the mass accretion rate $\dot{M}$ as:
\begin{align}
v_{r} = - \frac{\dot{M}}{2 \mathrm{\pi} r \Sigma} \  .
\label{eq:advection_v}
\end{align}

\subsection{Thermal Evolution}
\label{sec:thermal}

We calculate the time evolution of internal energy 
 $e_{\mathrm{in}}$ (per unit mass) in the flow 
by solving the energy equation,
\begin{align}
\frac{\mathrm{d} e_{\mathrm{in}}}{\mathrm{d}t} = -p\frac{\mathrm{d}}{\mathrm{d} t} \left(\frac{1}{\rho}\right)
+ \Gamma - \Lambda \  ,
\label{eq:thermal_1st}
\end{align}
where $p$ is the pressure, $\Gamma$ is the heating rate per unit mass and $\Lambda$ is the cooling rate per unit mass.
The internal energy and the temperature are related as 
\begin{align}
e_{\mathrm{in}} = \frac{1}{\gamma-1}\frac{k_{\mathrm{B}}T}{\mu m_{\mathrm{H}}} \  .
\label{eq:internal_ene}
\end{align}
The pressure is given by the ideal-gas equation of state,
\begin{align}
p = \frac{k_{\mathrm{B}} T}{\mu m_{\mathrm{H}}} \rho \  .
\label{eq:thermal_3}
\end{align}

We consider viscous heating, continuum cooling, $\mathrm{H}_{2}$ molecular and Li atomic line cooling and chemical cooling 
as the heating and cooling processes.  

The viscous heating rate per unit mass is
\begin{align}
\Gamma_{\mathrm{vis}} = \frac{9}{4} \nu \Omega^{2} \  ,
\label{eq:viscous_heat}
\end{align}
where $\nu$ is the kinematic viscosity.
For consistency with the assumption that the disc is steady and rotating at the Keplerian velocity, 
the kinematic viscosity must satisfy the relation 
\begin{align}
\nu = \frac{\dot{M}}{3 \mathrm{\pi} \Sigma} \  ,
\label{eq:kin_vis}
\end{align}
from the angular momentum conservation.
In our model, the kinematic viscosity is determined so that the surface density satisfies Equation~(\ref{eq:density}) 
for a given accretion rate.
The continuum radiation processes considered are $\mathrm{H}$ free-bound emission, $\mathrm{H}^{-}$ free-bound emission, $\mathrm{H}^{-}$ free-free emission, $\mathrm{H}$ free-free emission, $\mathrm{H}_{2}$-$\mathrm{H}_{2}$ collision induced emission (CIE) and $\mathrm{H}_{2}$-$\mathrm{He}$ CIE (reverse processes of 1-6 in Table~\ref{continuum}).
The continuum cooling rate in optically thin regime $\Lambda_{\mathrm{cont, thin}}$ is given by 
sum of Equation~(\ref{eq:thin_cool_f}) in Appendix~\ref{App:emissivity} over all the processes.  
Smoothly connecting the optically thin and thick limits, 
the continuum cooing rate in both regimes is given as \citep{Tanaka:2014}
\begin{align}
\Lambda_{\mathrm{cont}} = \frac{\Lambda_{\mathrm{cont, thin}}}
{1+ \tau_{\mathrm{P}} + \frac{3}{4} \tau_{\mathrm{R}} \tau_{\mathrm{P}} } \  ,
\label{eq:continuum}
\end{align}
where 
$\tau_{\mathrm{P(or \  R)}}$ is the Planck (Rosseland) mean optical depth, given by
\begin{align}
\tau_{\mathrm{P(or \  R)}} = \frac{1}{2} \Sigma \kappa_{\mathrm{P(or \  R)}} \  ,
\end{align}
with the Planck (Rosseland, respectively) mean opacity $\kappa_{\mathrm{P(R)}}$ calculated as in Appendix~\ref{App:opacity}. 
The $\mathrm{H}_{2}$-line cooling rate is given by multiplying the value in the optically thin regime 
$\Lambda_{\mathrm{H}_{2}, \mathrm{thin}}$ with the line-averaged escape probability $\overline{\beta}_{\mathrm{esc}}$:
\begin{align}
\Lambda_{\mathrm{H_{2}}} = \overline{\beta}_{\mathrm{esc}} \Lambda_{\mathrm{H_{2}, thin}}\mathrm{e}^{-\tau} \ , 
\end{align}
where $\tau=\sqrt{\tau_{\mathrm{P}}\tau_{\mathrm{R}}}$ is the effective optical depth for continuum radiation. 
We use fitting functions for $\overline{\beta}_{\mathrm{esc}}$ from \cite{Fukushima:2018} and 
for $\Lambda_{\mathrm{thin}, \mathrm{H}_{2}}$ from \cite{Glover:2015-11}, respectively.
As chemical cooling/heating processes, we consider those associated with H ionization/recombination 
and H$_{2}$ dissociation/formation. 
The Li cooling rate including line self-absorption effect 
$\overline{\beta}_{\mathrm{esc}}\Lambda_{\mathrm{Li,thin}}$ is given 
by Equation~\ref{Eq:Li_fit} (see Appendix~\ref{App:Li}). 
Considering also the continuum absorption, 
we obtain the Li cooling rate 
\begin{align}
\Lambda_{\mathrm{Li}} = \overline{\beta}_{\mathrm{esc}}\Lambda_{\mathrm{Li,thin}}\mathrm{e}^{-\tau}~.
\label{eq:Li_cooling}
\end{align}
The chemical cooling rate is
\begin{align}
\Lambda_{\mathrm{chem}} = \left( \chi_{\mathrm{H}}\frac{\mathrm{d}y(\mathrm{H^{+}})}{\mathrm{d}t} - 
\chi_{\mathrm{H}_{2}}\frac{\mathrm{d}y(\mathrm{H_{2}})}{\mathrm{d}t} \right)\frac{n_{\mathrm{H}}}{\rho} \  ,
\label{eq:chem_cool}
\end{align}
where $\chi_{\mathrm{H}}=13.6$ eV and $\chi_{\mathrm{H}_{2}}=4.48$ eV are the binding energies. 
The chemical fraction of species $i$, $y(i)$, is defined by the number fraction relative to hydrogen nuclei:
\begin{align}
y(i) = \frac{n(i)}{n_{\mathrm{H}}} \  ,
\label{eq:chem_fraction}
\end{align}
where $n(i)$ and $n_{\mathrm{H}}$ are the number densities of species $i$ and hydrogen nuclei, respectively. 
The latter is related to the mass density of the gas $\rho$ as
\begin{align}
n_{\mathrm{H}} = \frac{\rho}{(1+4y_{\rm He})m_{\mathrm{H}}}\ ,
\label{eq:number_density}
\end{align} 
where $y_{\mathrm{He}} = 8.333 \times 10^{-2}$ the  number fraction of He relative to hydrogen nuclei.

We do not include radiative heating by the central star in our calculation.
We estimated it under the assumption of optically-thin irradiation 
at the disc midplane and found it is generally lower than the viscous heating rate 
even at the Eddington luminosity because 
the coupling between the stellar radiation and disc is very inefficient 
due to the geometrical dilution of radiation ($\propto r^{-2}$) 
and transparency of the disc (typically $\tau \sim10^{-3}$-$10^{-4}$ in the case of atomic flow 
and $\sim10^{-6}$-$10^{-10}$ in the case of molecular flow at 1000 au, where 
the discs tend to be most gravitationally unstable; 
see Figs. \ref{atomic_mass}c and \ref{atomic_acc}c). 
The radiative heating does not change the disc structure
except in the outer regions in the case of low
accretion rate $\lesssim 10^{-2}~\mathrm{M}_{\odot}~\mathrm{yr}^{-1}$. 
In this case, the stellar radiation heating can be comparable to the viscous heating 
and raises the temperature to $\sim$1000 K. 
Although this somewhat stabilizes this part of the disc, the disc remains unstable 
(see Section~\ref{Sec:fragmentation_condition})
as in the case without stellar radiation. 
We thus expect that the effect of stellar radiation heating does not change our conclusion on
disc fragmentation.

\subsection{Chemical Reactions}
\label{sec:chemical}
We take into account the 20 chemical reactions
\footnote{We include the collisional dissociation of $\mathrm{H}_{2}$ by atomic helium impact (reaction 8 in Table~\ref{reactions}) 
for consistency with the expression used for the critical density of $\mathrm{H}_{2}$, where
the helium impact is also considered.}
among the following five H-bearing species,
$\mathrm{H}$, $\mathrm{H}_{2}$, $\mathrm{H}^{+}$, $\mathrm{H}^{-}$ and $\mathrm{e}$, 
summarized in Table~\ref{reactions} (see Appendix~\ref{App:inverse} for details).
All the reactions are paired with their reverse reactions.
We solve non-equilibrium kinetic equations for $\mathrm{H}$, $\mathrm{H}_{2}$, $\mathrm{H}^{+}$ and $\mathrm{e}$, 
while $\mathrm{H}^{-}$ fraction is calculated from the equilibrium between the forward and reverse reactions 
of reactions 2, 6 and 10 in Table~\ref{reactions}.
We assume $\mathrm{He}$ to be all neutral with fractional abundance $y_{\rm He}=8.333 \times 10^{-2}$ 
as our temperature range is limited below 10$^{4}$ K. 
We also assume that Li remains neutral with $y_{\rm Li}=4.15\times10^{-10}$ \citep{Mayer:2005} 
since they are still dominantly atomic in the calculated range 
and the small amount of Li ions does not affect thermal evolution at all. 
$\mathrm{H}_{2}$ photodissociation and $\mathrm{H}^{-}$ photodetachment by external radiation are not included. 
Because the number density is larger than $10^{8}~\mathrm{cm}^{-3}$ in our calculation range and the shielding of far-UV radiation is effective due to the large column density, 
the collisional dissociation works more efficiently 
than the dissociation due to external far-UV radiation. 
The effect of $\mathrm{H}^{-}$ photodetachment is also negligible because
$\mathrm{H_{2}}$ formation via the three body reaction is more effective
than the $\mathrm{H}^{-}$ channel at such high densities
as considered in this work. 

\subsection{Fragmentation Condition}
\label{Sec:fragmentation_condition}

We judge whether the disc fragments or not from the viscosity parameter
\begin{align}
\alpha = \frac{\nu \Omega}{c_{\mathrm{s}}^{2}} = \frac{G}{3 \mathrm{\pi}} \frac{\dot{M}}{c_{\mathrm{s}}^{3}} \  ,
\label{eq:alp_vis}
\end{align}
where we have used Equations~(\ref{eq:density}) and (\ref{eq:kin_vis}) for the last expression. 
We here adopt $\alpha > 1$ as the fragmentation condition following the results of 
\cite{Zhu:2012}, who claimed that discs fragment when the $\alpha$-value 
corresponding to the angular momentum transfer by gravitational torque exceeds unity
from two-dimensional numerical simulations of protoplanetary discs.
In other words, there is an upper limit on the angular momentum transported by the spiral arms and if 
more matter is accumulated than the disc can extract its angular momentum, the disc will fragment. 
In contrast, if $\alpha<1$ throughout, the disc remains stable and the angular momentum is transported 
by the gravitational torque.

\subsection{Model setup}
\label{Sec:setting}

We study cases with a range of the central stellar mass 
$10~\mathrm{M}_{\odot} \leq  M_{\ast} \leq 10^{5}~\mathrm{M}_{\odot}$ 
and the accretion rate 
$10^{-3}~\mathrm{M}_{\odot}~\mathrm{yr}^{-1} \leq  \dot{M} \leq 1~\mathrm{M}_{\odot}~\mathrm{yr}^{-1}$.
The disc structures at different central stellar masses can be regarded as temporal evolution 
around a growing SMS.
Recall that the typical accretion rate is $10^{-1}~\mathrm{M}_{\odot}~\mathrm{yr}^{-1}$ for SMS formation 
and $10^{-3}~\mathrm{M}_{\odot}~\mathrm{yr}^{-1}$ for ordinary first star formation, respectively. 
The calculated range of accretion rate encompasses these values. 

We set the outer boundary of the disc at 1000 au and assigning the temperature and chemical fractions 
there.
We evaluate the surface density by Equation~(\ref{eq:density}) using the temperature and the central stellar mass, 
and calculate the radial velocity using Equation~(\ref{eq:advection_v}).
Equation~(\ref{eq:kin_vis}) gives the kinematic viscosity that is required to maintain such a disc structure.
We then calculate the heating, cooling and chemical reaction rate, and update the temperature and the chemical fractions
along the inflow.
We continue the radial integration until the temperature exceeds $10^{4}~\mathrm{K}$, above which $\alpha < 0.4$ for $\dot{M} < 1~\mathrm{M}_{\odot}~\mathrm{yr}^{-1}$ (Equation~\ref{eq:alp_vis}) and thus the disc is always stable. 

We consider two different conditions in the flow at outer boundary: the atomic flow with  
temperature $T=3000$ K, molecular fraction $y({\rm H_2})=10^{-8}$, and ionization degree 
$y({\rm H^{+}})=10^{-6}$ and the molecular flow with $T=1000$ K, $y({\rm H_2})=0.495$, and $y({\rm H^{+}})=10^{-10}$.
Since ${\rm H^{-}}$ fraction is much lower than those of other species,   
the ${\rm H}$ and ${\rm e}$ fractions are calculated by 
$y(\mathrm{H}) = 1 - 2y(\mathrm{H}_{2}) - y(\mathrm{H}^{+}) \ $
and $y(\mathrm{H}^{+})=y(\mathrm{e})$, respectively, at the outer boundary. 
The atomic flow corresponds to an SMS forming in an atomically cooling cloud as a result of 
$\mathrm{H}_{2}$ photodissociation by external radiation, 
while the molecular one to that forming in an $\mathrm{H}_{2}$ cloud due, e.g., to baryonic streaming motions. 
For the former, we take the temperature and ionization degree from \cite{Omukai:2001} for the far-UV-irradiated cloud 
at number density $n_{\mathrm{H}}=10^{10}~\mathrm{cm}^{-3}$ with the far-UV intensity $J_{21} = 10^{5}$ 
(in unit of $10^{-21}$ erg $\mathrm{s^{-1}}~\mathrm{Hz^{-1}}~\mathrm{str^{-1}}~\mathrm{cm^{-2}}$)
and for the latter we take the self-regulated disc model from \cite{Schleicher:2016} at number density $n_{\mathrm{H}}=10^{10}~\mathrm{cm}^{-3}$.

\section{Result}
\label{Sec:result}

We present the disc structures and examine their gravitational stability
for atomic flows in Section~\ref{Sec:atomic_flow} and for molecular flows in Section~\ref{Sec:molecular_flow}, 
respectively.

\subsection{Atomic Accretion Flows}
\label{Sec:atomic_flow}

\subsubsection{Disc Structures}
\label{Sec:atomic_structure}

\begin{figure*}
 \begin{center}
 \begin{tabular}{c}
  {\includegraphics[scale=0.67]{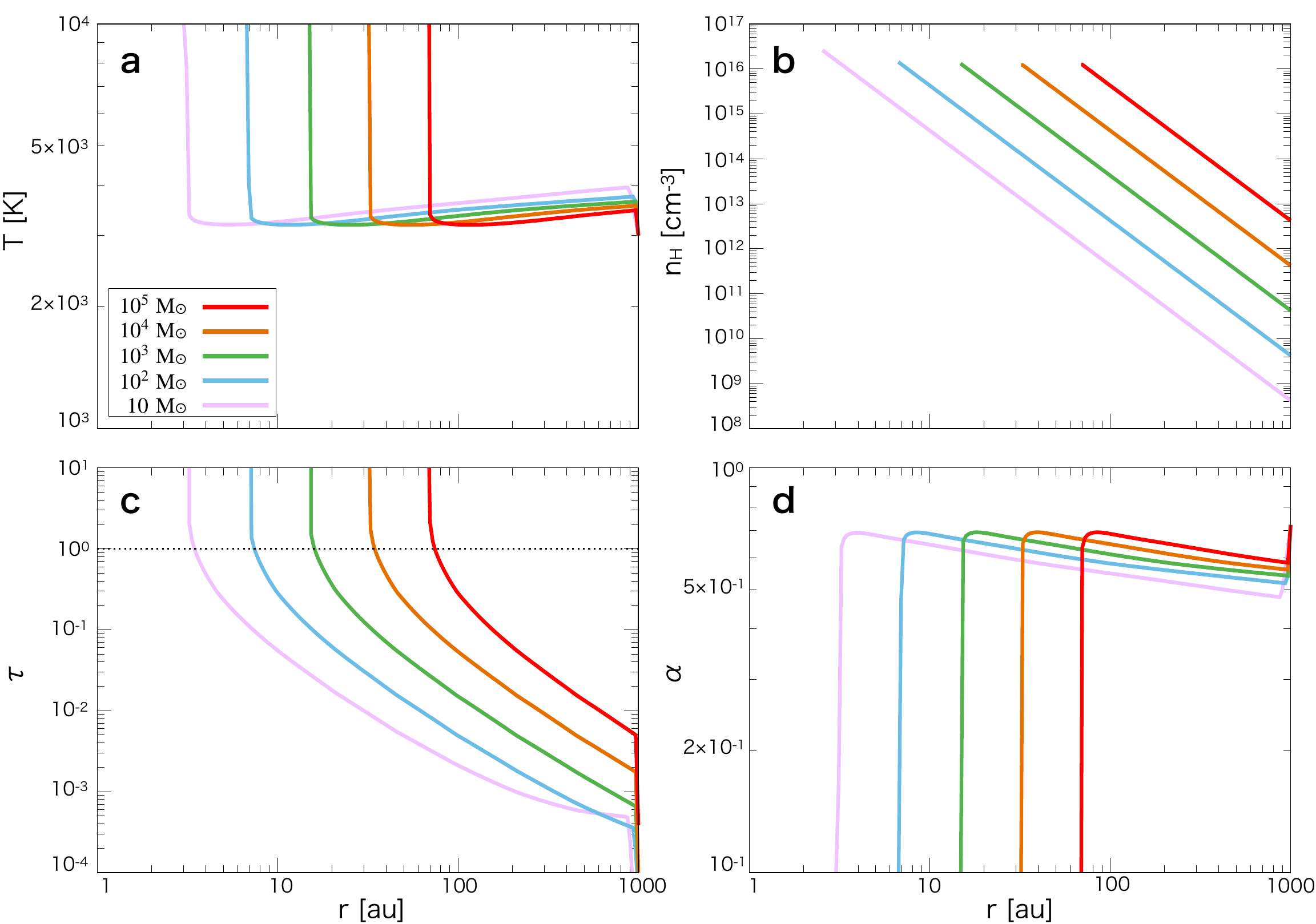}}
 \end{tabular}
 \caption{The disc structure for an atomic accretion flow 
with $\dot{M} = 10^{-1}~\mathrm{M}_{\odot}~\mathrm{yr}^{-1}$ at 
five different central stellar masses, $10^{5}~\mathrm{M}_{\odot}$ (red), $10^{4}~\mathrm{M}_{\odot}$
 (orange), $10^{3}~\mathrm{M}_{\odot}$ (green), $10^{2}~\mathrm{M}_{\odot}$ (blue) and
 $10~\mathrm{M}_{\odot}$ (purple). 
Panels show the radial distribution of 
(a) temperature, (b) number density, (c)
 effective optical depth for continuum radiation and (d) viscosity parameter $\alpha$.
 In panel (c), the black dotted line indicates $\tau=1$, which demarcates 
the optically thick and thin regimes.}
 \label{atomic_mass}
 \end{center}
\end{figure*}
\begin{figure*}
 \begin{center}
 \begin{tabular}{c}
  {\includegraphics[scale=0.67]{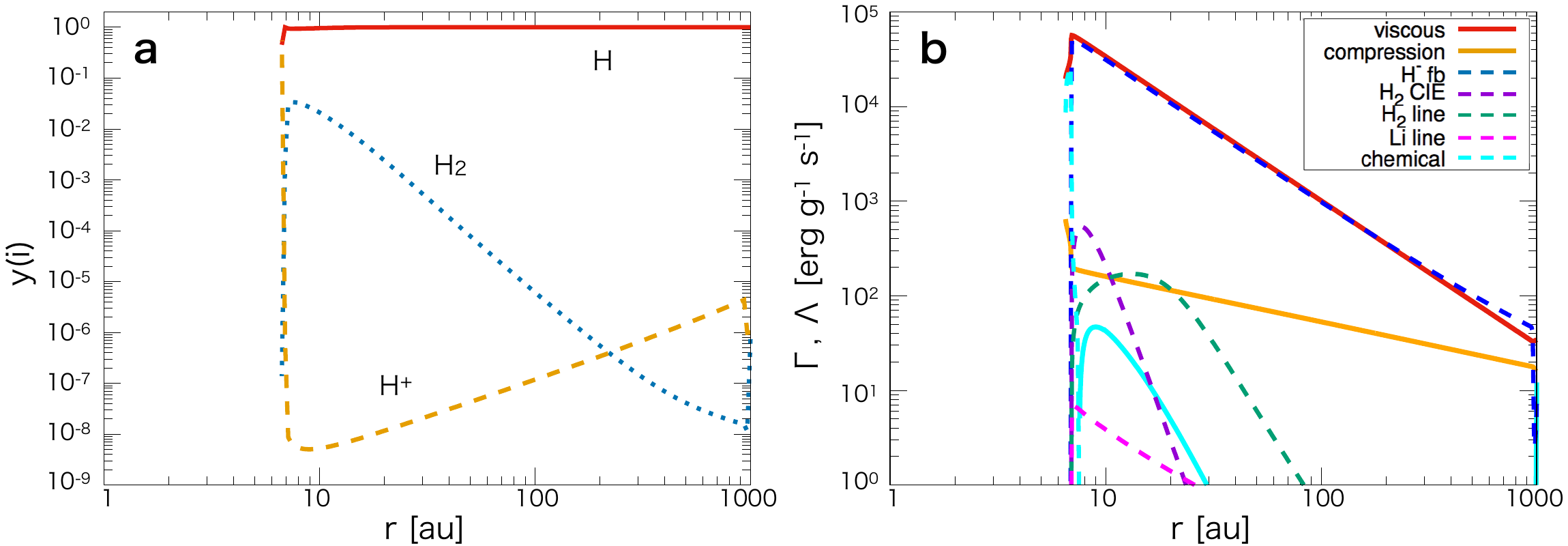}}
 \end{tabular}
 \caption{The radial distributions of (a) chemical fractions and (b)
 heating and cooling rates by individual processes 
 for the atomic accretion flow with $M_{\ast}
 = 10^{2}~\mathrm{M}_{\odot}$ and $\dot{M} = 10^{-1}~\mathrm{M}_{\odot}~\mathrm{yr}^{-1}$, 
 which are shown by blue lines in Fig.~\ref{atomic_mass}.  
 In the panel (a), the lines show the chemical
 fractions of $\mathrm{H}$ (red, solid), $\mathrm{H}_{2}$ (blue, dotted) and
 $\mathrm{H}^{+}$ (orange, dashed). 
 In the panel (b), solid and dashed lines correspond to heating and cooling rates, respectively. 
 The colours represent the
 viscous heating (red), compressional heating (orange), the $\mathrm{H}^{-}$
 free-bound emission cooling (blue), the $\mathrm{H}_{2}$ CIE cooling
 (purple), the $\mathrm{H}_{2}$ line emission cooling (green), 
 the Li line emission cooling (magenta) 
 and the chemical cooling and heating (cyan). 
 Cooling rates by other processes, i.e., $\mathrm{H}$ free-free
 emission, $\mathrm{H}^{-}$ free-free emission and 
$\mathrm{H}$ free-bound emission 
are much smaller and not shown in the figure.
}  \label{atomic_s2d-1}
 \end{center}
\end{figure*}
\begin{figure*}
 \begin{center}
 \begin{tabular}{c}
  {\includegraphics[scale=0.67]{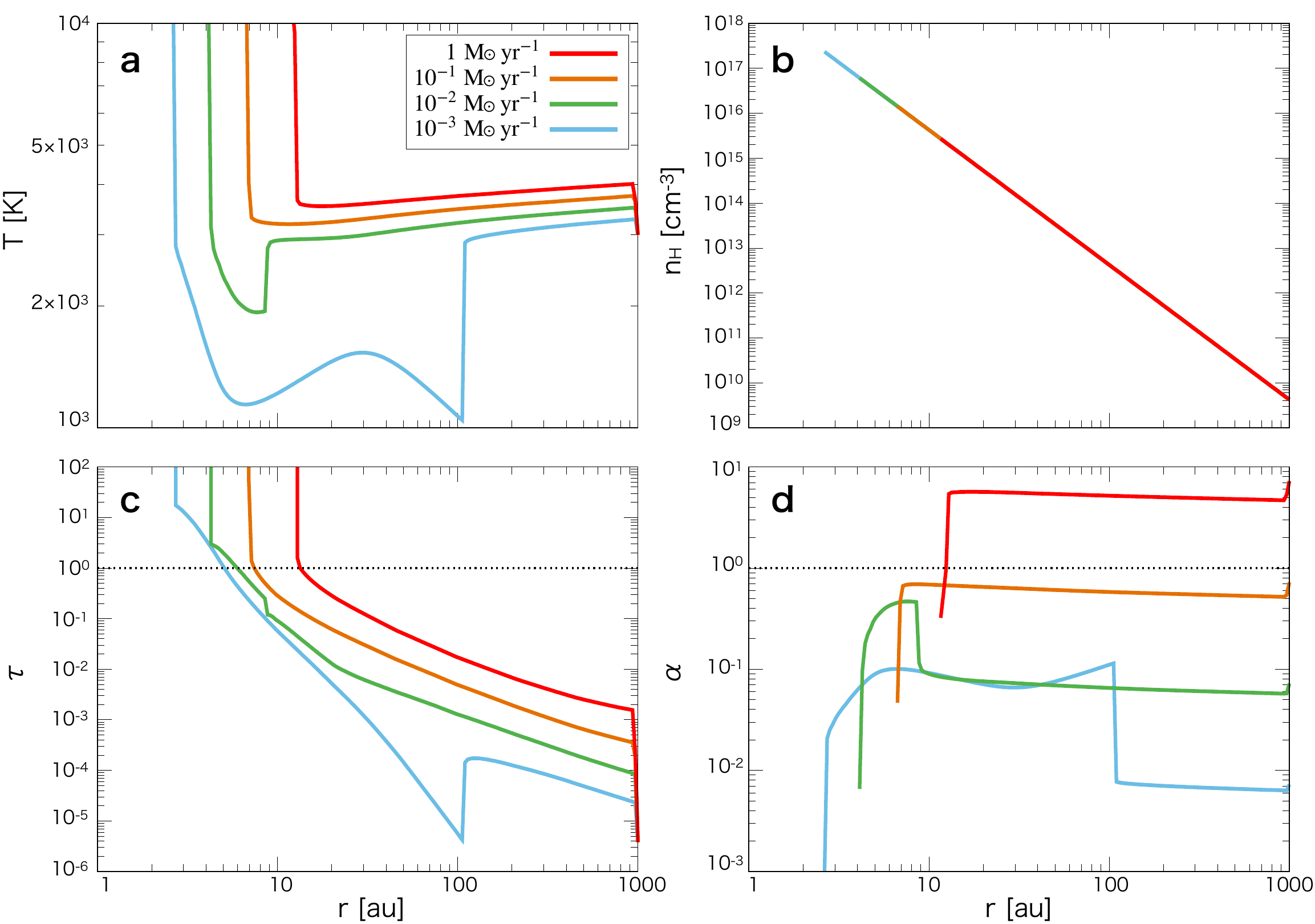}}
 \end{tabular}
 \caption{Same as Fig.~\ref{atomic_mass} but for 
atomic accretion flows with four different accretion rates 
 $1$ (red), $10^{-1}$ (orange), 
 $10^{-2}$ (green) and $10^{-3}~\mathrm{M}_{\odot}~\mathrm{yr}^{-1}$
 (blue) at the same stellar mass $M_{\ast} = 10^{2}~\mathrm{M}_{\odot}$. 
 In panel (d), the black dotted line shows $\alpha=1$, which
 demarcates the gravitationally stable/unstable regions} \label{atomic_acc}
 \end{center}
\end{figure*}
\begin{figure*}
 \begin{center}
 \begin{tabular}{c}
  {\includegraphics[scale=0.67]{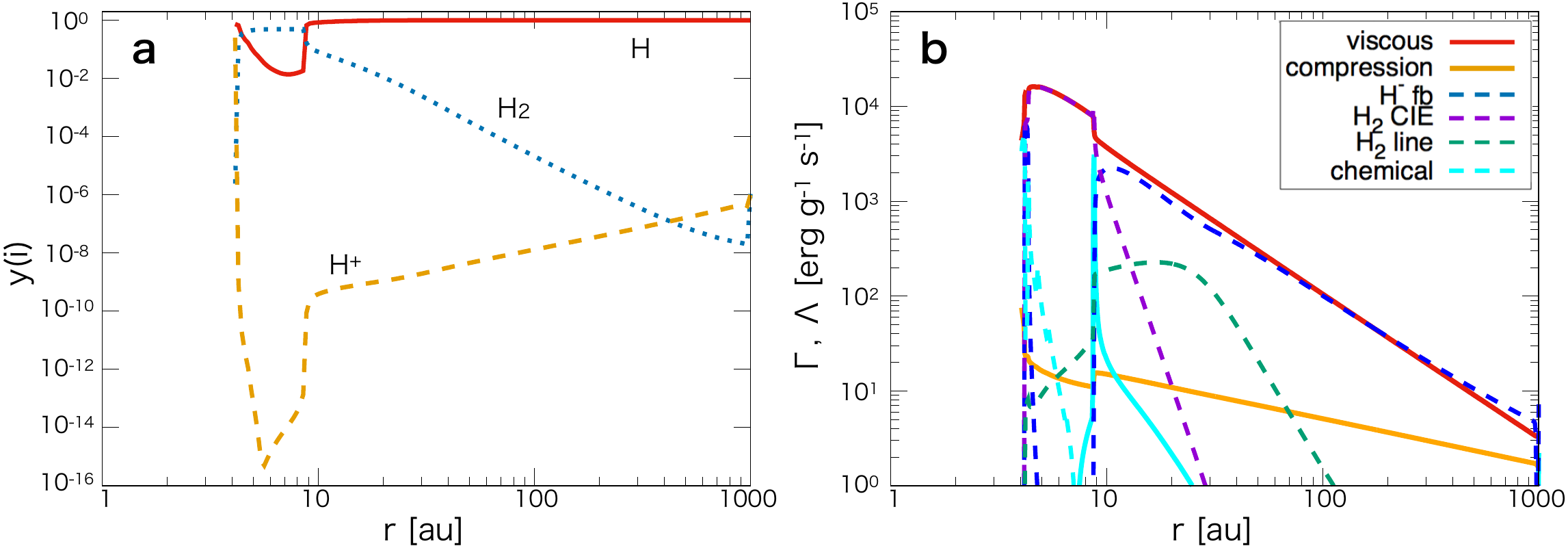}}
 \end{tabular}
 \caption{Same as Fig.~\ref{atomic_s2d-1},  
 but for the atomic accretion flow with $M_{\ast}
 = 10^{2}~\mathrm{M}_{\odot}$ and $\dot{M} = 10^{-2}~\mathrm{M}_{\odot}~\mathrm{yr}^{-1}$, 
 which are shown by green lines in Fig.~\ref{atomic_acc}.  
}  \label{atomic_s2d-2}
 \end{center}
\end{figure*}
\begin{figure}
 \begin{center}
 \begin{tabular}{c}
  {\includegraphics[scale=0.53]{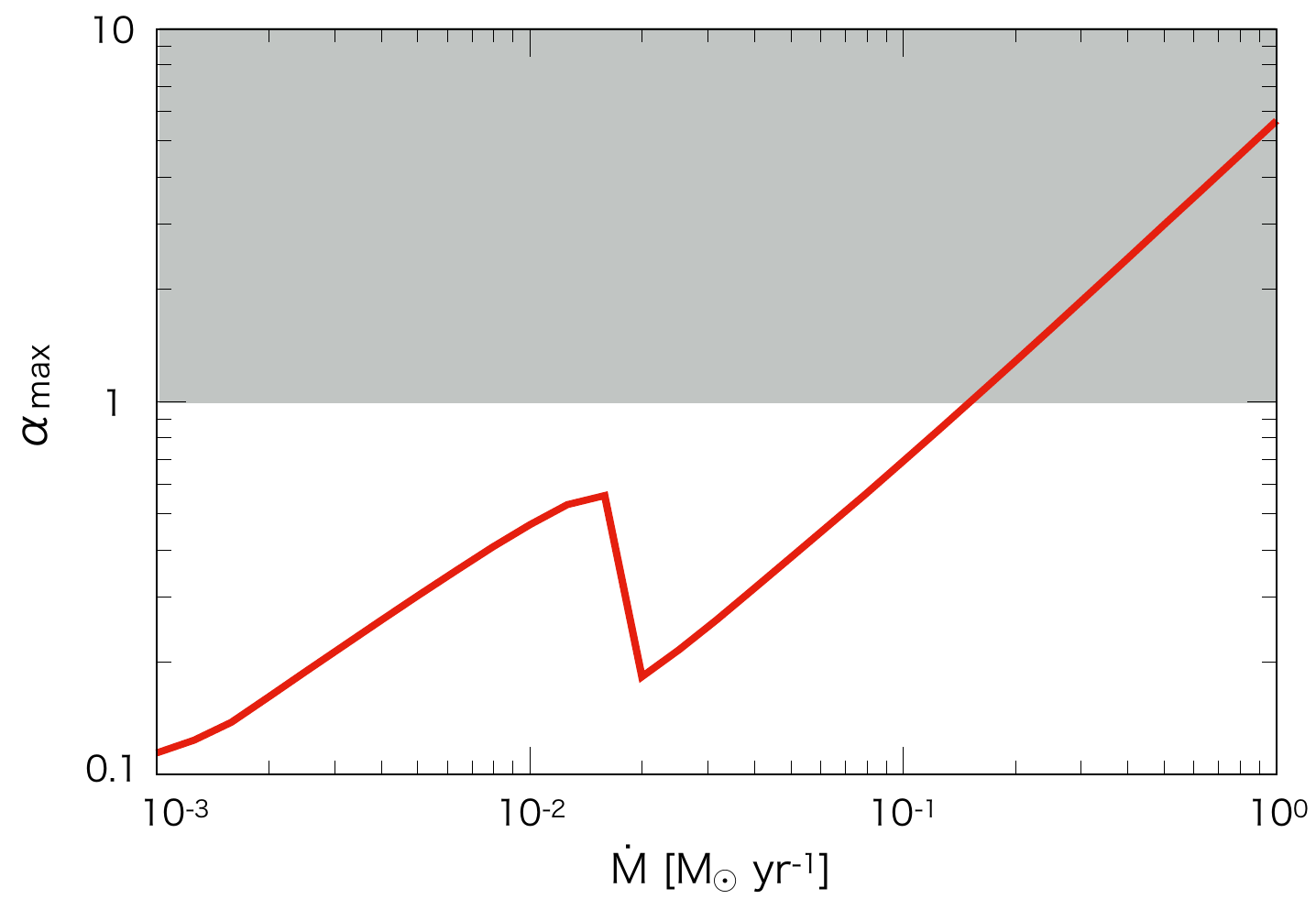}}
 \end{tabular}
 \caption{The maximum value of viscous parameter $\alpha$ in the disc 
as a function on the accretion rate for the atomic flows. 
 Here, we plot the result for $M_{\ast} = 10^{2}~\mathrm{M}_{\odot}$,  
 but those with other stellar masses are almost the same and 
 the lines overlap each other. 
 In the gray region, the maximum $\alpha$ value is
 larger than unity, implying that a part of the disc is gravitationally unstable.}  
\label{atomic_alpha_max}
 \end{center}
\end{figure}
The disc structures for atomic flows accreting at $\dot{M}=10^{-1}~\mathrm{M}_{\odot}~\mathrm{yr}^{-1}$ are shown in
Fig.~\ref{atomic_mass} for the central stellar mass of $M_{\ast}=10$,
$10^{2}$, $10^{3}$, $10^{4}$ and $10^{5}~\mathrm{M}_{\odot}$.  In all cases,  
the temperature first jumps up to around 4000 K at which the cooling and heating rates balances 
at the outer boundary at 1000 au, 
and then gradually decreases as the gas flows inward (Fig.~\ref{atomic_mass}a).  
When the optical depth exceeds 
unity, the radiative cooling becomes ineffective and the temperature 
suddenly rises to $>10^{4}$ K (Figs.~\ref{atomic_mass}a and c). 
Just before the optical depth reaches unity, the temperature takes its 
minimum at 3000 K, which does not depend on the central stellar mass. 
The density depends on the radius as $n_{\mathrm{H}} \propto r^{-3}$ (Fig.~\ref{atomic_mass}b), 
as can be derived analytically from Equations~(\ref{eq:angular_velocity}), (\ref{eq:density}), (\ref{eq:mass_density}) and (\ref{eq:scale_height}). 

To understand the temperature evolution in Fig.~\ref{atomic_mass}a, 
we show in Fig.~\ref{atomic_s2d-1} the radial distribution of the chemical fractions (Fig.~\ref{atomic_s2d-1}a) 
and the heating and the cooling rates (Fig.~\ref{atomic_s2d-1}b) for the case with $M_{\ast} = 10^{2}~\mathrm{M}_{\odot}$ 
and $\dot{M} = 10^{-1}~\mathrm{M}_{\odot}~\mathrm{yr}^{-1}$ (corresponding to the blue lines in Fig.~\ref{atomic_mass}).
Dominant chemical species is atomic $\mathrm{H}$ from 1000 au to 10 au, where the 
ionization starts and the temperature increases (Fig.~\ref{atomic_s2d-1}a), 
and 
dominant heating and cooling processes are the viscous heating and the $\mathrm{H}^{-}$ free-bound emission cooling, 
respectively, and the temperature is determined by their balance (Fig.~\ref{atomic_s2d-1}b). 
This is also true for all the cases shown in Fig.~\ref{atomic_mass}.
Note that the contribution of lithium line cooling is always negligible 
although \cite{Mayer:2005} insisted lithium cooling can be important, 
if it is at LTE value,  
at the relevant density and temperature ranges (see their Fig. 4).
This is because the actual rate is far smaller than the LTE rate 
as the electron number density is generally several orders of 
magnitude below the critical density, even with the photon trapping effect. 
In the inner region, although the electron density increases,
the continuum optical depth exceeds unity before reaching the critical density 
and the cooling rate decreases exponentially thereafter.

The minimum temperature is set by the balance between 
the viscous heating (from Equations~\ref{eq:scale_height}, \ref{eq:mass_density} and \ref{eq:kin_vis}), 
\begin{align}
\Gamma_{\mathrm{vis}} = \frac{3}{8\mathrm{\pi}} \frac{\Omega^{3}}{\rho c_{\mathrm{s}}} \dot{M} \ .
\label{eq:viscous_heat2}
\end{align}
and 
the cooling rate by $\mathrm{H}^{-}$ free-bound emission 
(from Equations~\ref{eq:continuum} and \ref{eq:thin_cool_f}) 
\begin{align}
\Lambda_{\mathrm{H}^{-} \mathrm{fb}} 
&= \frac{4\mathrm{\pi} f_{\mathrm{H}^{-} \mathrm{fb}}(T) y(\mathrm{e}) n_{\mathrm{H}}^{2}}{\rho} \notag \\
&= \frac{4 \mathrm{\pi} \Omega^{3} f_{\mathrm{H}^{-} \mathrm{fb}}(T) K(T)^{-\frac{1}{2}}} 
{\left[ 2\mathrm{\pi} (1+4y_{\mathrm{He}})m_{\mathrm{H}} G \right]^{\frac{3}{2}} \rho} \  , 
\label{eq:Hmfb_cool}
\end{align} 
where $f_{\mathrm{H}^{-} \mathrm{fb}}(T)$ is a function of temperature (see Equation~\ref{eq:thin_cool_f}) and 
we have assumed that mostly atomic gas of $y(\mathrm{H}) \sim 1$ (Fig.~\ref{atomic_s2d-1}a) with 
small ionization degree given by the equilibrium value 
\begin{align}
y(\mathrm{e}) = 1/\sqrt{ n_{\mathrm{H}} K(T) } \  ,
\label{eq:ye_saha}
\end{align}
due to high enough density ($\sim 10^{16}~\mathrm{cm}^{-3}$) 
at the temperature minimum (Fig.~\ref{atomic_mass}a and b), 
where the equilibrium constant $K(T)$ is again a function only of temperature (see Appendix~\ref{App:inverse}).
We use the optically thin rate  
as the optical depth is still small at this epoch (Fig.~\ref{atomic_mass}c).
Equating Equations~(\ref{eq:viscous_heat2}) and (\ref{eq:Hmfb_cool}), we obtain
\begin{align}
f_{\mathrm{H}^{-} \mathrm{fb}}(T)K(T)^{-\frac{1}{2}} c_{\mathrm{s}} \propto \dot{M} \ , 
\label{eq:vis_Hmfb}
\end{align}
which implicitly gives the temperature $T$ as a function of $\dot{M}$.
For example, for $\dot{M} = 10^{-1} \  \mathrm{M}_{\odot}~\mathrm{yr}^{-1}$, the equilibrium temperature is $T=$3100 K, 
consistent with that in Fig.~\ref{atomic_mass}a.
The slight outward temperature increase in Fig.~\ref{atomic_mass}
is due to the smaller actual cooling rate than that 
by Equation~(\ref{eq:Hmfb_cool}) as a result of smaller electron 
fraction than the equilibrium value used above. 
It should be noted that the minimum temperature in the disc given as a solution of Equation (\ref{eq:vis_Hmfb}) 
depends only on the accretion rate and not on the central stellar mass, as seen in Fig.~\ref{atomic_mass}a.

Next we study the dependence on the accretion rate.
The disc structures for the central stellar mass $M_{\ast}=10^{2}~\mathrm{M}_{\odot}$ are shown in Fig.~\ref{atomic_acc}
for four different accretion rates $\dot{M} = 10^{-3}$, $10^{-2}$, $10^{-1}$ and $1~\mathrm{M}_{\odot}~\mathrm{yr}^{-1}$.
The temperature profiles in the high accretion rate cases of $10^{-1}$ and $1~\mathrm{M}_{\odot}~\mathrm{yr}^{-1}$ behave 
in a similar way as seen before in Fig.~\ref{atomic_mass}:
the gradual inward decrease from 4000 K to 3000 K followed by the sudden jump when the disc becomes optically thick.
At lower accretion rates, i.e., $10^{-3}$ and $10^{-2}~\mathrm{M}_{\odot}~\mathrm{yr}^{-1}$, however, 
the temperature experiences a steep drop to 1000-2000K at some radius before eventually jumping up to $>10^{4}$ K.
This is a result of H$_2$ formation and its CIE cooling.  
For illustration, we show the radial distribution of the chemical fractions (Fig.~\ref{atomic_s2d-2}a), 
and the heating and cooling rates (Fig.~\ref{atomic_s2d-2}b) for the case of 
$\dot{M} = 10^{-2}~\mathrm{M}_{\odot}~\mathrm{yr}^{-1}$ (green lines in Fig.~\ref{atomic_acc}).
At radii outer than 9 au, the gases are mostly atomic and 
the temperatures are set by the balance between 
the viscous heating and the cooling by the $\mathrm{H}^{-}$ free-bound emission.
Once the temperature becomes as low as 3000 K, transition to the molecular phase occurs, accompanied with 
a sudden temperature drop to 2000K due to the resultant $\mathrm{H}_{2}$ CIE cooling.
The transition radius moves outward to 100 au in the case of $\dot{M} = 10^{-3}~\mathrm{M}_{\odot}~\mathrm{yr}^{-1}$
due to lower viscous heating and temperature.
Note that this molecular transition in the disc has not been recognized before \citep{Inayoshi:2014-12, Latif:2015}
and only becomes able to be followed by our detailed thermal and chemical modeling. 
Similar dependence of disc structures on the accretion rate is also observed for other central stellar masses.
\begin{figure*}
 \begin{center}
 \begin{tabular}{c}
  {\includegraphics[scale=0.67]{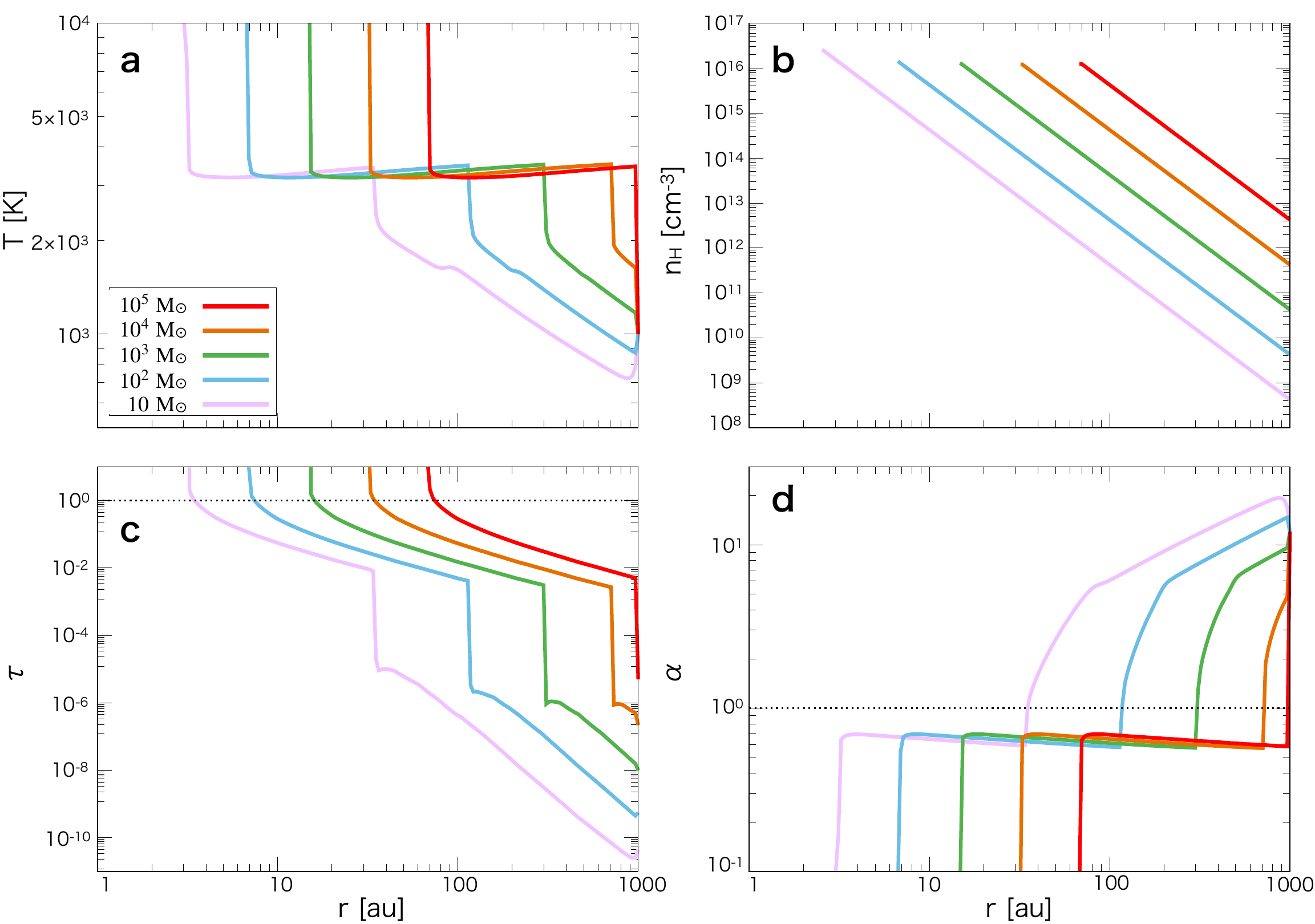}}
 \end{tabular}
 \caption{Same as Fig.~\ref{atomic_mass} 
for a molecular accretion flow 
with $\dot{M} = 10^{-1}~\mathrm{M}_{\odot}~\mathrm{yr}^{-1}$ at 
five different central stellar masses, $10^{5}~\mathrm{M}_{\odot}$ (red), $10^{4}~\mathrm{M}_{\odot}$
 (orange), $10^{3}~\mathrm{M}_{\odot}$ (green), 
 $10^{2}~\mathrm{M}_{\odot}$ (blue) and
 $10~\mathrm{M}_{\odot}$ (purple).}
 \label{molecular_mass}
 \end{center}
\end{figure*}
\begin{figure*}
 \begin{center}
 \begin{tabular}{c}
  {\includegraphics[scale=0.67]{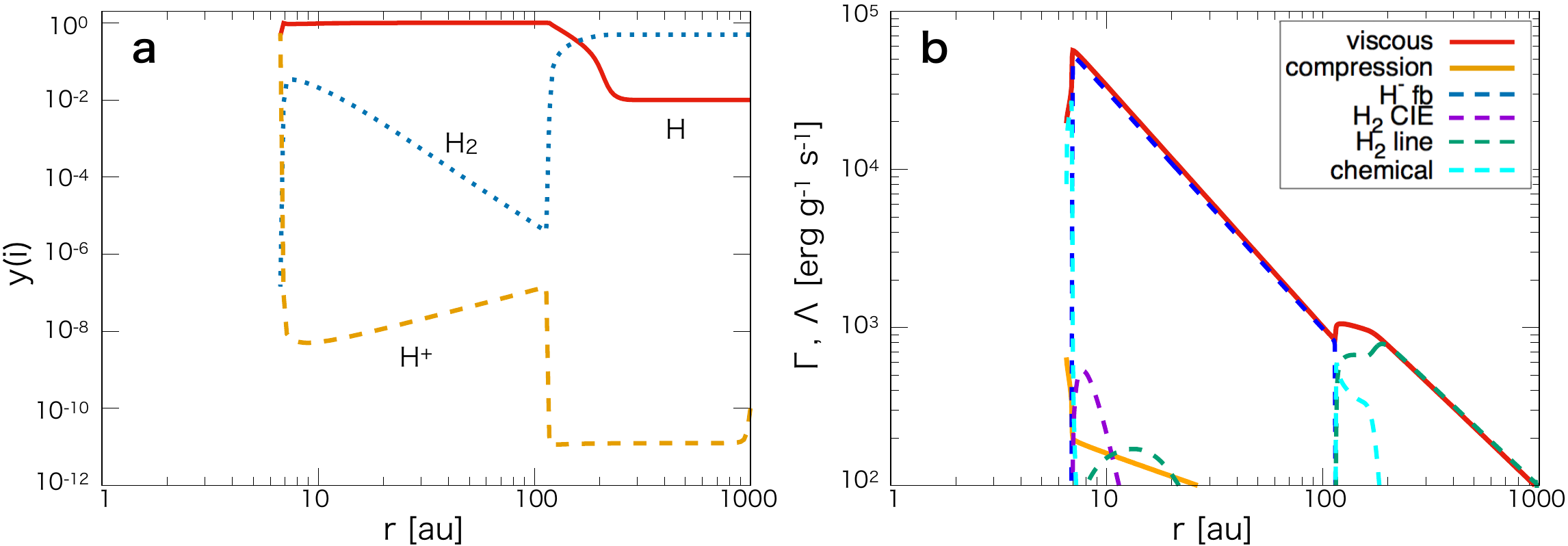}}
 \end{tabular}
 \caption{
 Same as Fig.~\ref{atomic_s2d-1},  
 but for the molecular accretion flow with $M_{\ast}
 = 10^{2}~\mathrm{M}_{\odot}$ and $\dot{M} = 10^{-1}~\mathrm{M}_{\odot}~\mathrm{yr}^{-1}$, 
which are shown by blue lines in Fig.~\ref{molecular_mass}.  
}
 \label{molecular_s2d-1}
 \end{center}
\end{figure*}
\begin{figure*}
 \begin{center}
 \begin{tabular}{c}
  {\includegraphics[scale=0.67]{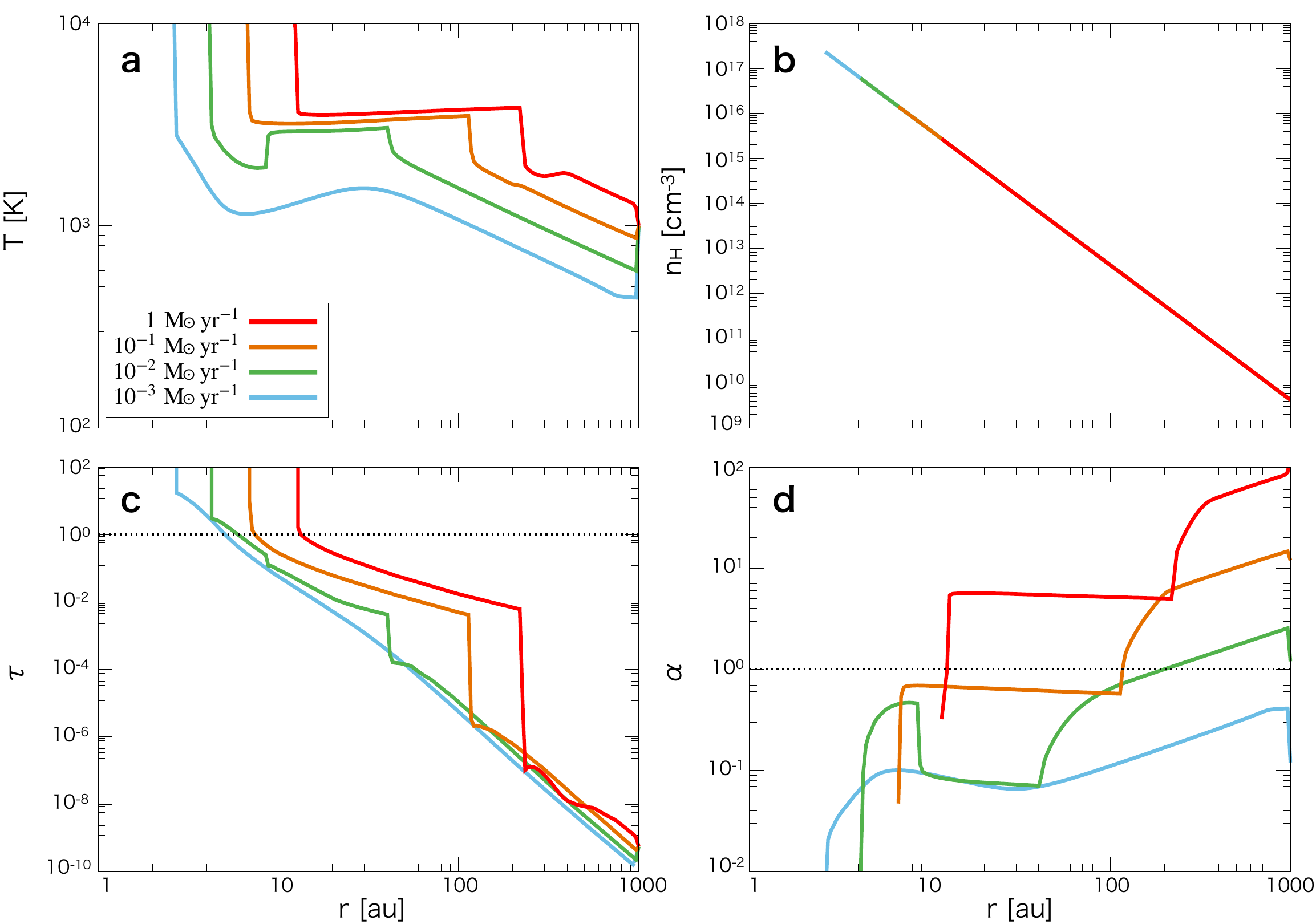}}
 \end{tabular}
 \caption{Same as Fig.~\ref{atomic_acc} but for molecular accretion
 flows with four different accretion rates 
$1$ (red), $10^{-1}$ (orange), 
$10^{-2}$ (green) and $10^{-3}~\mathrm{M}_{\odot}~\mathrm{yr}^{-1}$ (blue) 
at the same $M_{\ast} = 10^{2}~\mathrm{M}_{\odot}$.} 
 \label{molecular_acc}
 \end{center}
\end{figure*}
\begin{figure*}
 \begin{center}
 \begin{tabular}{c}
  {\includegraphics[scale=0.67]{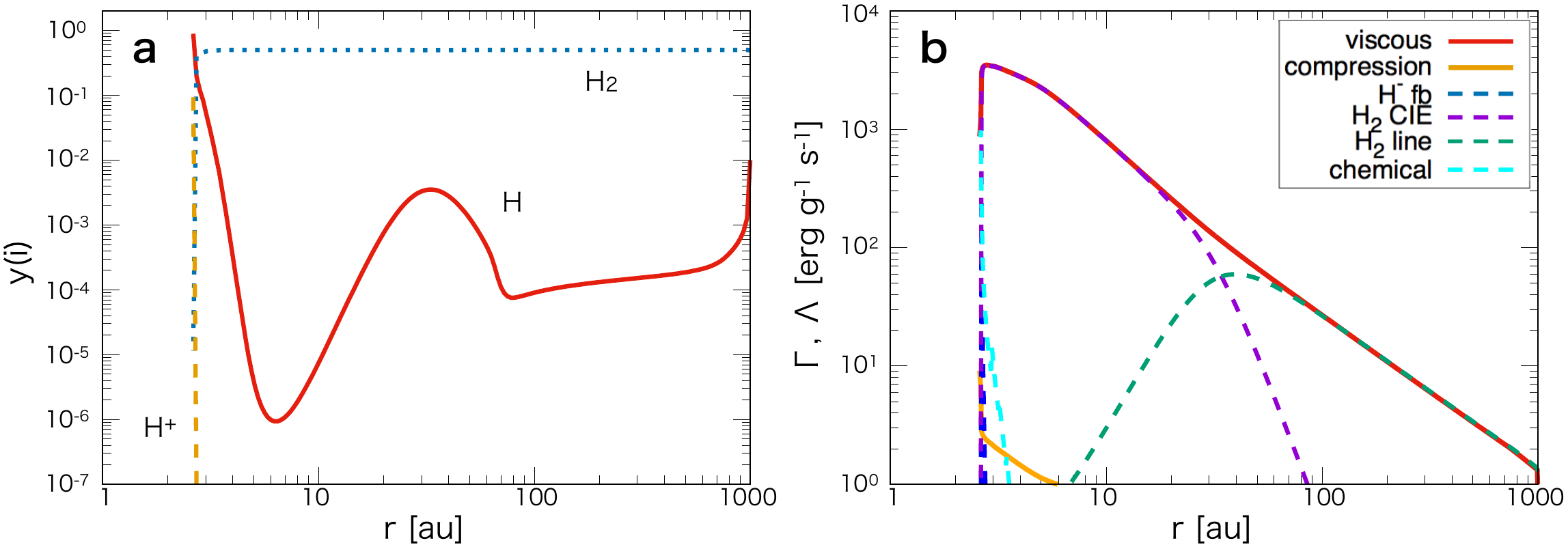}}
 \end{tabular}
 \caption{Same as Fig.~\ref{atomic_s2d-1} but for the molecular accretion flow with $M_{\ast} = 10^{2}~\mathrm{M}_{\odot}$ 
and $\dot{M} = 10^{-3}~\mathrm{M}_{\odot}~\mathrm{yr}^{-1}$, which are shown by blue line in Fig.~\ref{molecular_acc}.}
 \label{molecular_s2d-3}
 \end{center}
\end{figure*}

\subsubsection{Gravitational Instability of Discs}
\label{Sec:atomic_fragmentation}

We now examine the disc instability for the atomic inflow 
using the $\alpha$ parameters. 
The radial distributions of the $\alpha$-value 
are shown for different stellar masses but
with the same accretion rate of $\dot{M}=10^{-1}~\mathrm{M}_{\odot}~\mathrm{yr}^{-1}$ in Fig.~\ref{atomic_mass}d, 
while those for the same mass $M_{\ast}=10^{2}~\mathrm{M}_{\odot}$ but 
with four different accretion rates are shown in Fig.~\ref{atomic_acc}d.
Since $\alpha \propto \dot{M}/T^{3/2}$ (Equation~\ref{eq:alp_vis}) for constant $\mu$, 
the radial variation of $\alpha$ follows that of the temperature but 
in a upside-down way.
As seen in Fig.~\ref{atomic_acc}d, 
$\alpha$ tends to have higher value for higher accretion rate.  
With high accretion rate of $\dot{M} \geq 10^{-1}~\mathrm{M}_{\odot}~\mathrm{yr}^{-1}$, 
$\alpha$ is almost constant radially in the optically thin part of the disc, 
while with lower rate $\dot{M} \leq 10^{-2}~\mathrm{M}_{\odot}~\mathrm{yr}^{-1}$, 
the maximum value of $\alpha$ is reached at the the temperature minimum by the runaway $\mathrm{H}_{2}$ formation.

As seen in Section~\ref{Sec:atomic_structure}, the minimum temperature depends only on the accretion rate and not on the central stellar mass, 
so does the maximum value of $\alpha$, $\alpha_{\mathrm{max}}$ (see Fig.~\ref{atomic_mass}d).
Fig.~\ref{atomic_alpha_max} shows $\alpha_{\mathrm{max}}$ as a function of the accretion rate.
There are two branches where $\alpha_{\mathrm{max}} \propto \dot{M}$ below and above 
$\dot{M} = 2 \times 10^{-2}~\mathrm{M}_{\odot}~\mathrm{yr}^{-1}$, which correspond, respectively, to 
the case where $\alpha_{\mathrm{max}}$ is attained in the $\mathrm{H}_{2}$ phase (as shown in Fig.~\ref{atomic_s2d-2})
and the case where the gas always remains atomic in the optically thin part (Fig.~\ref{atomic_s2d-1}).
Here we adopt $\alpha > 1$ as the condition for disc fragmentation (Section~\ref{Sec:fragmentation_condition}).
As seen in Fig.~\ref{atomic_alpha_max}, this occurs for $\dot{M} \ga 10^{-1}~\mathrm{M}_{\odot}~\mathrm{yr}^{-1}$. 
This critical accretion rate for disc fragmentation can be obtained by
substituting $\alpha = 1$ and the sound velocity for neutral gas with the minimum temperature of 3000 K (Fig.~\ref{atomic_mass}a)
into Equation~(\ref{eq:alp_vis}):
\begin{align}
\dot{M}_{\mathrm{crit}} = 0.1~\mathrm{M}_{\odot}~\mathrm{yr}^{-1} \left( \frac{T}{3000~\mathrm{K}} \right)^{\frac{3}{2}} \  .
\end{align}
Since the typical accretion rate for SMSs formation falls on around this critical value, 
whether fragmentation actually occurs or not would depend on detailed dynamics 
of collapse and accretion (see Section~\ref{Sec:discussion_conclusion}). 

\subsection{Molecular Accretion Flows}
\label{Sec:molecular_flow}

\subsubsection{Disc Structures}
\label{Sec:molecular_structure}

The disc structure for a molecular flow is shown in Fig.~\ref{molecular_mass} 
in the case with accretion rate $\dot{M}=10^{-1}~\mathrm{M}_{\odot}~\mathrm{yr}^{-1}$
for different central stellar masses 
$M_{\ast}=10$, $10^{2}$, $10^{3}$, $10^{4}$ and $10^{5}~\mathrm{M}_{\odot}$.
Soon after the beginning of calculation, 
the temperature near the outer boundary quickly converses to the thermal equilibrium value, 
which is higher for the higher accretion rate (Fig.~\ref{molecular_mass}a).
In the highest mass case of $M_{\ast}= 10^{5}~\mathrm{M}_{\odot}$, this temperature reaches $\sim 3000$ K and 
the gas becomes atomic immediately as a result of collisional H$_2$ dissociation.
At lower stellar masses $M_{\ast} \leq 10^{4}~\mathrm{M}_{\odot}$, the initial temperature at the outer boundary is 
lower than $\sim 2000$ K but increases as the gas flows inward, first gradually up to $\sim 2000$ K and 
then suddenly to $\sim 3000$ K.
As an example, for the case of $M_{\ast} = 10^{2}~\mathrm{M}_{\odot}$ (blue lines in Fig.~\ref{molecular_mass}), 
radial profiles of the chemical fractions (Fig.~\ref{molecular_s2d-1}a) 
and the heating and cooling rates (Fig.~\ref{molecular_s2d-1}b) are 
presented in Fig.~\ref{molecular_s2d-1}. 
At 100 au, transition from the molecular to atomic phases proceeds due to the H$_{2}$ collisional dissociation 
(Fig.~\ref{molecular_s2d-1}a) and simultaneously the dominant cooling process shifts from the $\mathrm{H}_{2}$-line emission to 
the $\mathrm{H}^{-}$ free-bound emission (Fig.~\ref{molecular_s2d-1}b), which causes the abrupt temperature increase to 3000 K.
After the transition to $\mathrm{H}$, the temperature is set by the balance between 
the viscous heating and the $\mathrm{H}^{-}$ free-bound emission cooling and 
slightly decreases inward until the disc becomes optically thick, 
as already seen for the atomic flows (Fig.~\ref{atomic_mass}a).
Note that the decrease of $\mathrm{H}_{2}$ fraction with density here looks more abrupt 
(Figs.~\ref{molecular_mass}a and~\ref{molecular_s2d-1}a) 
than in the case of collapsing prestellar cores \cite[e.g.,][]{Omukai:2001}. 
This difference can be attributable to smaller chemical cooling rate in our case:
the latent heat associated with $\mathrm{H}_{2}$ dissociation is extracted 
in a longer timescale for density enhancement, i.e., disc viscous time, than 
the free-fall timescale of collapsing cores, and thus at a smaller chamical cooling rate. 
The temperature in the disc increases more easily, resulting in the sudden $\mathrm{H}_{2}$ dissociation.

Flows with different accretion rates $\dot{M}=10^{-3}$, $10^{-2}$, $10^{-1}$ and $1 \  \mathrm{M}_{\odot}~\mathrm{yr}^{-1}$ 
are shown in Fig.~\ref{molecular_acc} for the same stellar mass $M_{\ast}=10^{2}~\mathrm{M}_{\odot}$. 
The dependence of the disc structure on the accretion rate is similar also in the cases with other values of stellar mass.
In high accretion-rate cases of $10^{-1}$ and $1~\mathrm{M}_{\odot}~\mathrm{yr}^{-1}$, 
the disc structures are the same as discussed above in Fig.~\ref{molecular_mass}:
the temperature suddenly increases from 2000 K to 3000 K at atomic phase transition, then stays at 3000 K in the optically thin 
atomic gas and eventually jumps up to $>10^{4}$ K when the disc becomes optically thick.
With lower accretion rate of $\dot{M} = 10^{-2}~\mathrm{M}_{\odot}~\mathrm{yr}^{-1}$, 
the gas returns to molecular phase again at 10 au and temperature drops to 2000K
as seen for the atomic accretion flow with low accretion rates (Fig.~\ref{atomic_acc}).
In the case of $\dot{M} = 10^{-3}~\mathrm{M}_{\odot}~\mathrm{yr}^{-1}$,  
the temperature variation is more gradual than in the other cases.
The chemical fractions and the heating and cooling rates in this case 
are shown in Fig.~\ref{molecular_s2d-3}.
Unlike in higher accretion rate cases, 
$\mathrm{H}_{2}$ now survives as the dominant species until the flow reaches the innermost part. 
The $\mathrm{H}$ fraction increases from 70 au.
After peaking at 30 au, it then decreases inward as the temperature also decreases (Fig.~\ref{molecular_acc}).
The temperature decrease at $<30$ au is caused by the cooling by $\mathrm{H}_{2}$ CIE (Fig.~\ref{molecular_s2d-3}b).
The temperature more inside is set by the thermal balance between this cooling and the viscous heating 
until the disc becomes optically thick to the $\mathrm{H}_{2}$ collision-induced absorption. 

\subsubsection{Disc Fragmentation}
\label{Sec:molecular_fragmentation}

\begin{figure}
 \begin{center}
 \begin{tabular}{c}
  {\includegraphics[scale=0.53]{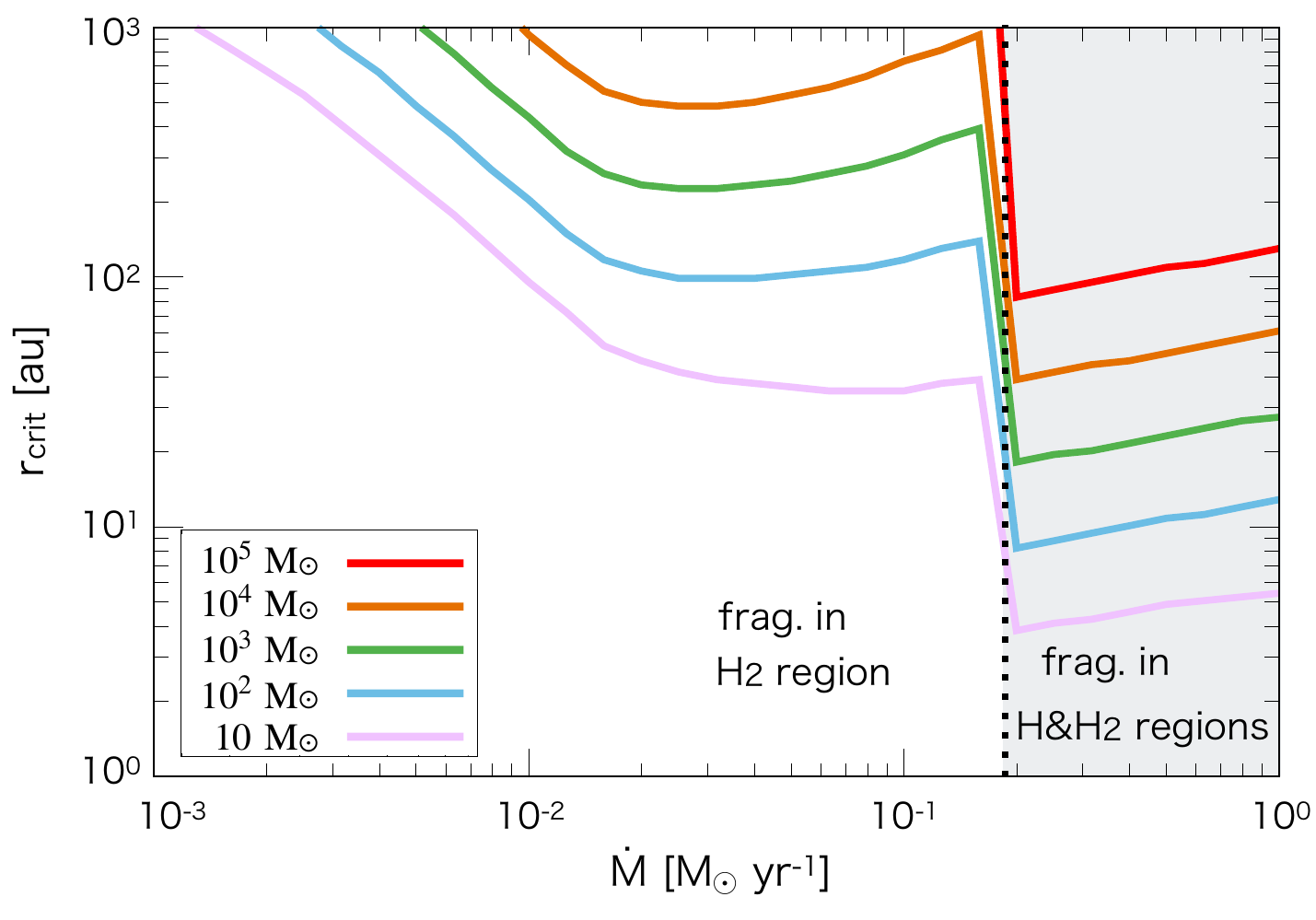}}
 \end{tabular}
 \caption{The critical radius for gravitational instability for 
 molecular accretion flows as a function of the accretion rate. 
 The different colours show the different
 central stellar masses: $10^{5}~\mathrm{M}_{\odot}$ (red), $10^{4}~\mathrm{M}_{\odot}$
 (orange), $10^{3}~\mathrm{M}_{\odot}$ (green), $10^{2}~\mathrm{M}_{\odot}$ (blue) and
 $10~\mathrm{M}_{\odot}$ (purple). The disc is gravitationally unstable both in the
 outer molecular and inner atomic regions in the case with $\dot{M}
 \gtrsim 2\times 10^{-1}~\mathrm{M}_{\odot}~\mathrm{yr}^{-1}$ (shaded
 region), while the instability occurs only in the outer molecular 
 region otherwise.}
 \label{critical_radius}
 \end{center}
\end{figure}
Next, we examine the disc fragmentation condition for the molecular accretion flows.
The radial distributions of $\alpha$-value for such flows are presented 
 in Fig.~\ref{molecular_mass}d for $\dot{M} = 10^{-1}~\mathrm{M}_{\odot}~\mathrm{yr}^{-1}$ at different stellar masses.
The $\alpha$-value takes its maximum just inside the outer boundary where the temperature is the lowest, 
except in the case with $M_{\ast} = 10^{5}~\mathrm{M}_{\odot}$ (see Section~\ref{Sec:molecular_structure} for details).
Going inward, the $\alpha$-value decreases along with the increasing temperature. 
The $\mathrm{H}_{2}$ dissociation and resultant temperature jump to 3000 K causes the corresponding drop of 
$\alpha$ below unity in all the cases.
We thus expect that fragmentation, if any, occurs only in the outer $\mathrm{H}_{2}$ region.
The cases for different accretion rates are shown for $M_{\ast} = 10^{2}~\mathrm{M}_{\odot}$
in Fig.~\ref{molecular_acc}d.
Again, the maximum $\alpha$ is reached around the outer boundary and its value 
is almost proportional to the accretion rate (Equation~\ref{eq:alp_vis}) as in the case of atomic flows.

As discussed above, 
for the molecular flows, the $\alpha$-value tends to be higher at outer radius, 
suggesting that if the disc is large enough the $\alpha$ exceeds unity at some radius.
We can thus define the critical radius here as the radius where $\alpha = 1$ is satisfied 
and plot it in Fig.~\ref{critical_radius}
as a function of the accretion rate for each central stellar mass.
If the size of a disc is smaller than this, the disc remains stable. 
Otherwise, its outer part would become gravitationally unstable and fragment. 
In all the cases shown in Fig.~\ref{critical_radius}, the critical radius becomes abruptly small 
at $\sim 10^{-1}~\mathrm{M}_{\odot}~\mathrm{yr}^{-1}$ 
because the condition of $\alpha<1$ is attained only when the gas ionizes at innermost radius for higher accretion rate.
Also, the critical radius increases with the central stellar mass, 
reflecting that the temperature profiles shift toward outer radius as seen in Fig.~\ref{molecular_mass}a.
With increasing stellar mass by accretion, the disc at a given radius becomes more stable and 
only the more outer part can be gravitationally unstable. 

\section{Summary and Discussion}
\label{Sec:discussion_conclusion}
Seed black hole (BH) formation by the direct collapse and its subsequent growth is one viable scenario 
for the supermassive black hole (SMBH) formation in the early Universe.
In this framework, a small protostar is first formed by the gravitational collapse of a cloud, 
and grows to supermassive star (SMS) by rapid accretion, and finally the SMS collapses by general 
relativistic instability into a seed BH.
In this paper, 
we have investigated gravitational instability of the discs around rapidly accreting protostars 
with mass $10-10^{5}~\mathrm{M}_{\odot}$ at rate $10^{-3}-1~\mathrm{M}_{\odot}~\mathrm{yr}^{-1}$ 
by way of the steady $\alpha$-disc model with detailed treatment of chemical and thermal processes. 
We have considered two possible compositions of the inflow, i.e., mostly atomic and molecular gases,
at the disc outer boundary.
We have constructed marginally gravitationally stable disc structure. 
By comparing the required value of viscosity parameter $\alpha$ to maintain such structure and the possible maximum value of 
unity, we have judged the gravitational instability of discs. 

In the case of atomic flows, we have found the following two types of disc structures depending on the accretion rate.
At high rate ($\gtrsim 10^{-2}~\mathrm{M}_{\odot}~\mathrm{yr}^{-1}$), 
the gas composition remains atomic in the optically thin part with temperature of $\sim 3000$ K, 
set by the H$^-$ free-bound emission.
At lower rate, in contrast, the compositional transition to the H$_2$ phase occurs at some radius 
and $\mathrm{H}_{2}$ collision-induced emission dominates the cooling further inward. 
In both cases, once the disc becomes optically thick, the temperature jumps up to $> 10^4$ K and the gas is ionized. 
The discs are gravitationally unstable if the accretion rate is 
higher than $10^{-1}~\mathrm{M}_{\odot}~\mathrm{yr}^{-1}$ (Fig.~\ref{atomic_alpha_max}) regardless of the central stellar mass.

If the inflow is mainly composed of a molecular gas, 
the $\mathrm{H}_{2}$-line cooling keeps the temperature below 2000 K in outer part of the disc 
although the temperature gradually increases inward. 
With the accretion rate higher than $10^{-2}~\mathrm{M}_{\odot}~\mathrm{yr}^{-1}$, 
the $\mathrm{H}_{2}$ dissociation occurs while the disc is still optically thin and the 
temperature jumps up to 3000K.  
With a lower rate, the gas remains in the molecular form until it becomes optically thick to 
the H$_{2}$ collision-induced absorption. 
For a given accretion rate and stellar mass, such disc is always unstable at sufficiently large radius (Fig.~\ref{critical_radius}).
This critical radius is $10^{2}-10^{3}$ au for $\lesssim 10^{-1}~\mathrm{M}_{\odot}~\mathrm{yr}^{-1}$ 
depending on the stellar mass, but decreases abruptly by an order of magnitude with higher accretion rates.

~

Based on the analysis above, let us
briefly discuss the possibility of disc fragmentation during SMS formation
in different formation sites, whose main
component can be either atomic or molecular.
First, for a star forming from the atomic gas, 
the accretion rate can be estimated as
\begin{align}
\dot{M} = \phi \frac{c_{\mathrm{s,col}}^{3}}{G} = 0.1\ \phi \left(\frac{c_{\mathrm{s,col}}}{8\ \mathrm{km\ s^{-1}}}\right)^{3} \ \mathrm{M}_{\odot} \  \mathrm{yr}^{-1}\  .
\end{align}
where $c_{\mathrm{s,col}}$ is the effective sound speed in the collapsing cloud 
including also the effect of turbulence, magnetic field etc., 
and $\phi$ is a factor that reflects the manner of collapse: $\phi\sim 10$ for dynamically collapsing cores
and $\phi\sim 1$ for quasi-statically collapsing cores
\citep{Hunter:1977}.  According to numerical simulations for the collapse induced by atomic cooling 
\citep[e.g.,][]{Inayoshi:2014-11}, 
$c_{\mathrm{s,col}}$ is $\sim 10~\mathrm{km\ s^{-1}}$, which indicates 
the occurrence of disc fragmentation largely depends on detailed manner of the collapse.
Second, for the molecular-gas case, where SMSs are supposed to form in
sites with large streaming velocity between baryon and dark matter, 
whether the size of the discs is bigger or smaller than the critical one 
(Fig.~\ref{critical_radius}) matters.  
In \cite{Hirano:2017}'s numerical simulations, disc radii are
$\sim$1000 au in most of the cases, while the
accretion rates fluctuate between $10^{-1}~\mathrm{M}_{\odot}~\mathrm{yr}^{-1}$ and 
$10^{-2}~\mathrm{M}_{\odot}~\mathrm{yr}^{-1}$. 
Thus, the discs are larger than the critical radius and expected to 
fragment before the central stellar mass reaches $ \sim 10^{4}~\mathrm{M}_{\odot}$ (Fig.~\ref{critical_radius}), 
which is actually the case as found by \cite{Hirano:2017}.

If a disc fragments into multiple clumps, the accretion onto the central protostar
becomes episodic with repeating bursts accompanying falls of the clumps 
and intervening quiescent phases. With a quiescent phase longer than 
the Kelvin-Helmholtz timescale, 
the protostar contracts to a main-sequence star and starts emitting ionization photons, 
which can terminate further growth of the central star
\citep{Sakurai:2015,Sakurai:2016}.  
For quantitative modeling of the accretion-rate variability, hydrodynamical simulations 
incorporating with the chemical and thermal networks are awaited.

The outcome of the disc fragmentation studied here might be supermassive binary stars, 
which can be detectable by future gravitational wave observations \citep{Chon:2018}. 
For example, if an equal mass binary system with $10^{4}~\mathrm{M}_{\odot}$ forms and 
ends up with the merger of the remnant BHs with the
same mass, the emitted gravitational waves would be detectable by 
{\it Deci-hertz Interferometer Gravitational wave Observatory} ({\it DECIGO}: \citealt{Kawamura:2011}) 
and {\it Laser Interferometer Space Antenna} ({\it LISA}: \citealt{Amaro-Seoane:2012}) up to redshift
$z \sim 30$ and $\sim 10$, respectively.

In this work, we have adopted the critical $\alpha$-value of unity 
as the condition for disc fragmentation in light of numerical 
simulations of protoplanetary discs \citep{Zhu:2012}. 
Fragmentation condition is, however, still in dispute 
\citep[cf.,][]{Gammie:2001,Meru:2012,Rice:2014,Takahashi:2016}. 
Furthermore,
it is not clear whether the condition obtained for protoplanetary discs
is applicable to the those around SMSs. 
If the critical value of $\alpha$ becomes smaller, 
discs will fragment more easily.
In particular, for atomic flows with low accretion rate
($\lesssim 10^{-2}~\mathrm{M}_{\odot}~\mathrm{yr}^{-1}$), the cold inner regions 
by $\mathrm{H}_{2}$ cooling (see
Figs.~\ref{atomic_acc} and \ref{atomic_s2d-2}) may also become gravitationally unstable 
if a smaller critical $\alpha$-value is adopted.

Here the disc structure is obtained under the thin disc approximation. 
To check this, we calculated the aspect ratio $H/r$ for all the cases considered in this paper 
and found that in most cases it falls in the range $\sim10^{-1}-10^{-2}$, justifying this approximation.  
Owing to the dependence $H/r\propto r^{1/2}M_{\ast}^{-1/2}$, however, the thin disc approximation 
becomes worse for lower $M_{\ast}$ and larger $r$. 
In fact, the ratio $H/r$ becomes $\sim$~1 in our smallest mass case considered $M_{\ast}=10~\mathrm{M}_{\odot}$ 
around the outer edge of the disc at $\sim$ 1000 au.
In reality, however, 
the disc size would be smaller than our adopted value of 1000 au in such an early stage
since the disc size increases with time accompanying the accretion of large angular momentum material 
originally located at the outer part of the parental dense core. 

Considering uncertainties above, further studies are needed to conclude 
whether SMSs actually form or not.
We plan to perform hydrodynamics simulations that follow the
formation of SMSs from the collapse of cores, where the disc
fragmentation and subsequent evolution of fragments supposedly play an
important role.


\section*{Acknowledgments}
The authors would like to thank Kohei Inayoshi, Daisuke Nakauchi, 
Hidekazu Tanaka and Hidenobu Yajima for discussions.  We also thank
Shingo Hirano and Takashi Hosokawa for providing their simulation
results and the anonymous reviewer for useful comments on improving 
thermal and chemical modelling.  RM acknowledges financial support from the Graduate Program on
Physics for the Universe of Tohoku University. This work is supported in
part by MEXT/JSPS KAKENHI Grant Number 17H06360 (KS and KO)
and 17H01102, 17H02869 (KO) and by
NAOJ ALMA Scientific Research Grant Numbers 2016-02A (ST).



\appendix

\section{Radiative processes}

\subsection{Continuum cooling rate in the optically thin regime}
\label{App:emissivity}

For continuum emission cooling, we consider the reverse of absorption
processes in Table~\ref{continuum}. 
Here, all the emission processes are in the form either 
of $\mathrm{A+B \to C+\gamma}$ or $\mathrm{A+B \to
C+D+\gamma}$, and their emissivities $\eta_{\nu}$ are proportional to the
product of $n(\mathrm{A})$ and $n(\mathrm{B})$.
More specifically, the emissivity of photons per unit volume via
free-bound emission ($\mathrm{A + e \to B + \gamma}$) is given by \citep{Omukai:2001}: 
\begin{align}
\eta_{\nu} = \frac{2h \nu^{3}}{c^{2}} \frac{z_{\mathrm{B}}}{z_{\mathrm{A}} z_{\mathrm{e}}} &\left( \frac{m_{\mathrm{B}}h^{2}}{2 \mathrm{\pi} m_{\mathrm{A}} m_{\mathrm{e}} k_{\mathrm{B}} T} \right)^{\frac{3}{2}} \sigma_{\nu} \notag \\
&\times \mathrm{exp} \left[ - \frac{h \nu - \chi_{B}}{k_{\mathrm{B}} T} \right] n(A) n(\mathrm{e}) \  ;
\label{eq:eta_fb}
\end{align}
that via free-free emission ($\mathrm{A} + \mathrm{e} \to \mathrm{A} + \mathrm{e} + \gamma$) or 
$\mathrm{H}_{2}$ CIE ($\mathrm{H}_{2}(=A) + \mathrm{B} \to \mathrm{H}_{2} + \mathrm{B} + \gamma$) 
\begin{align}
\eta_{\nu} &= \frac{2h \nu^{3}}{c^{2}} \sigma_{\nu} \  \mathrm{exp} \left[ - \frac{h \nu}{k_{\mathrm{B}} T} \right] n(A) \  .
\label{eq:eta_ff}
\end{align}
By integrating over frequency, the cooling rate by $i$-th process in the optically thin regime is 
\begin{align}
\Lambda_{i, {\rm thin}}=\frac{4 \mathrm{\pi}}{\rho} \int \eta_{i, \nu} \mathrm{d} \nu = \frac{4 \mathrm{\pi}}{\rho} f_{i} (T) n(\mathrm{A}) n(\mathrm{B}) \  ,
\label{eq:thin_cool_f}
\end{align}
with $f_{i}(T)$ a function only of the temperature.  

The coefficients $f_{i}(T)$ are calculated by numerically integrating Equations~(\ref{eq:eta_fb}) and (\ref{eq:eta_ff}) 
in the range $10^{-3}\,\mathrm{eV} < h \nu < 10^{2}\,\mathrm{eV}$ and the results are fitted for the temperature range
$500\,\mathrm{K}< T < 20000\,\mathrm{K}$ except in the case of
$\mathrm{H}^{-}$ free-free emission ($900\,\mathrm{K} < T <
20000\,\mathrm{K}$ in this case) in the form of
\begin{align}
\mathrm{log}~f_{\mathrm{i}}(T) = &a_{0} + a_{1} \  \mathrm{log}~T + a_{2} \  (\mathrm{log}~T)^{2} + a_{3} \  (\mathrm{log}~T)^{3} \notag \\
&+ a_{4} \  (\mathrm{log}~T)^{4} \  .
\label{eq:fitting_func}
\end{align}
The fitting constants $a_{0}$, $a_{1}$, $a_{2}$, $a_{3}$ and $a_{4}$ are presented in Table~\ref{fit_emission} 
for each emission process.  
The error in the fitting is generally less than a few percent. 

\begin{table*}
 \begin{center}
 \caption{Continuum absorption processes}
 \label{continuum}
  \scalebox{1.0}[1.1]{ 
  \begin{tabular}{c l l c} \hline \hline
    Number & Name & Process &  Reference \\ \hline
    1 & $\mathrm{H}$ bound-free & $\mathrm{H} + \gamma \to \mathrm{H}^{+} + \mathrm{e}$ &  {\small \cite{Rybicki:1979}} \\
    2 & $\mathrm{H}^{-}$ bound-free & $\mathrm{H}^{-} + \gamma \to \mathrm{H} + \mathrm{e}$ & {\small \cite{John:1988}}\\
    3 & $\mathrm{H}^{-}$ free-free & $\mathrm{H} + \mathrm{e} + \gamma \to \mathrm{H} + \mathrm{e}$ & {\small \cite{John:1988}} \\
    4 & $\mathrm{H}$ free-free & $\mathrm{H}^{+} + \mathrm{e} + \gamma \to \mathrm{H}^{+} + \mathrm{e}$ & {\small \cite{John:1988}} \\
    5 & $\mathrm{H}_{2}$-$\mathrm{H}_{2}$~CIA & $\mathrm{H}_{2}$(v,J)$ + \mathrm{H}_{2} + \gamma \to \mathrm{H}_{2}$(v',J')$ + \mathrm{H}_{2}$ & {\small \cite{Borysow:1997}}\\
    6 & $\mathrm{H}_{2}$-$\mathrm{He}$~CIA & $\mathrm{H}_{2}$(v,J)$ + \mathrm{He} + \gamma \to \mathrm{H}_{2}$(v',J')$ + \mathrm{He}$ & {\small \cite{Borysow:1997}}\\
    7 & $\mathrm{H}$ Rayleigh & $\mathrm{H} + \gamma \to \mathrm{H} + \gamma'$ & {\small \cite{Kurucz:1970}}\\
    8 & $\mathrm{H}_{2}$ Rayleigh & $\mathrm{H}_{2} + \gamma \to \mathrm{H}_{2} + \gamma'$ & {\small \cite{Dalgarno:1962}} \\
    9 & Thomson & $\mathrm{e} + \gamma \to \mathrm{e} + \gamma'$ & {\small \cite{Rybicki:1979}} \\ \hline
  \end{tabular}
  }
 \end{center}
\end{table*}
\begin{table*}
 \begin{center}
 \caption{Coefficients for fitting function (\ref{eq:fitting_func})}
 \label{fit_emission}
  \scalebox{1.2}[1.3]{ 
  \begin{tabular}{l|c c c c c} \hline \hline
    process & $a_{0}$ & $a_{1}$ & $a_{2}$ & $a_{3}$ & $a_{4}$ \\ \hline
    $\mathrm{H}$ ~free-bound & -22.410 & -0.51833 & 2.3058d-3 & 0.0000 & 0.0000 \\ 
    $\mathrm{H}^{-}$~free-bound & -31.211 & 1.1678 & -3.0473d-2 & 0.0000 & 0.0000 \\ 
    $\mathrm{H}^{-}$~free-free & -38.564 & 3.4681 & -0.17880 & 0.0000 & 0.0000 \\ 
    $\mathrm{H}$~free-free & -28.030 & 0.54235 & -5.3291d-3 & 0.0000 & 0.0000 \\ 
    $\mathrm{H}_{2}$-$\mathrm{H}_{2}$~CIE & 75.232 & -1.4487d+2 & 65.126 & -12.304 & 0.84166 \\ 
    $\mathrm{H}_{2}$-$\mathrm{He}$~CIE & 87.458 & -1.6412d+2 & 75.157 & -14.482 & 1.0120 \\ \hline
  \end{tabular}
  }
 \end{center}
\end{table*}

\subsection{Planck and Rosseland mean opacities}
\label{App:opacity}

The cooling rate in the optically thick regime is reduced from 
the optically thin value due to the photon trapping effect, 
which is taken into account as in Equation~(\ref{eq:continuum}) by using 
the Planck and Rosseland optical depths ($\tau_{\rm P}$ and $\tau_{\rm R}$). 
Here we present the method for calculation of Planck and Rosseland mean 
opacities. 
The continuum absorption processes taken into account are listed in Table~\ref{continuum}.  
In addition to those considered in \cite{Omukai:2001}, 
we also consider the $\mathrm{H}_{2}$ Rayleigh scattering following \citealt{Mayer:2005}. 
The opacity $\kappa_\nu$ per unit mass of gas via a pure absorption process, i.e., 
bound-free absorption ($\mathrm{A + \gamma \to B + \mathrm{e}}$), 
free-free absorption ($\mathrm{A + \mathrm{e} + \gamma \to A + \mathrm{e}}$)
and $\mathrm{H}_{2}$ collision-induced absorption (CIA; $\mathrm{H}_{2}(={\rm A}) + \mathrm{B} + \gamma \to \mathrm{H}_{2} + \mathrm{B}$), 
is given by considering stimulated emission: 
\begin{align}
\kappa_\nu = \frac{\sigma_{\nu} n(A)}{\rho} \left( 1 - \mathrm{exp} \left( - \frac{h\nu}{k_{\mathrm{B}} T} \right) \right),
\label{eq:abs-bf}
\end{align}
with the Planck constant $h$ and the photon frequency $\nu$. 
That for scattering ($\mathrm{A + \gamma \to A + \gamma'}$) is  
\begin{align}
\kappa_\nu = \frac{\sigma_{\nu} n(A)}{\rho} \  .
\label{eq:abs-sca}
\end{align}
Note that the cross-sections $\sigma_{\nu}$ for free-free absorption and $\mathrm{H}_{2}$ CIA are proportional
to the density of collision partners, $n(\mathrm{e})$ and $n(\mathrm{B})$, respectively.

The Planck mean opacity can be calculated by
\begin{align}
\kappa_{\mathrm{P}} = {\displaystyle \sum_{i=1}^6 \int \kappa_{\mathrm{i},\nu} B_{\nu} \mathrm{d}\nu}
\bigg/{\displaystyle\int B_{\nu} \mathrm{d}\nu} \  ,
\label{eq:abs-Planck}
\end{align}
where subscript $i$ represents an individual process in Table~\ref{continuum} and the sum is over only pure absorption processes,  
while the Rosseland mean opacity is 
\begin{align}
\frac{1}{\kappa_{\mathrm{R}}}  = 
{\displaystyle \sum_{\mathrm{i}=1}^9 \int \kappa_{\mathrm{i},\nu}^{-1}\frac{\partial B_{\nu}}{\partial T} \mathrm{d}\nu}
\bigg/{\displaystyle\int \frac{\partial B_{\nu}}{\partial T} \mathrm{d}\nu} \ , 
\label{eq:abs-Rosseland}
\end{align}
where the sum is over all the absorption and scattering processes.

The photon trapping effect becomes remarkable only when the disc is nearly optically thick. 
In such condition, the chemical fractions are approximately in the equilibrium values, 
which are given by Saha-Boltzmann equations. 
For example, among species {\rm A}, {\rm B}, and {\rm C} for which a reaction $\mathrm{A \leftrightarrow B + C}$ exists,
the equilibrium fractions are written as
\begin{align}
\left[ \frac{n(\mathrm{A})}{n(\mathrm{B}) n(\mathrm{C})} \right]^{*} = \frac{z_\mathrm{A}}{z_\mathrm{B} z_\mathrm{C}} 
\left( \frac{m_\mathrm{A}}{m_\mathrm{B} m_\mathrm{C}} \right)^{\frac{3}{2}} &\left( \frac{2 \mathrm{\pi} k_{\mathrm{B}} T}{h^{2}} \right)^{-\frac{3}{2}} 
\mathrm{exp} \left( -\frac{\Delta\chi}{k_{\mathrm{B}} T} \right) \  ,
\label{eq:chemSaha}
\end{align}
where the superscript $*$ means the chemical equilibrium value,
$\Delta\chi=\chi_\mathrm{B}+\chi_\mathrm{C}-\chi_\mathrm{A}$ is the
difference in chemical binding energy, and
$m_\mathrm{X}$,  $z_\mathrm{X}$ and $\chi_\mathrm{X}$ are
the mass, partition function and chemical potential energy of species $\mathrm{X}$, respectively.
We adopt $z_{\mathrm{H}} = z_{\mathrm{e}} = 2$, $z_{\mathrm{H}^{+}} =
z_{\mathrm{H}^{-}} = 1$, $\chi_{\mathrm{H}} = 13.6 \ \mathrm{eV}$,
$\chi_{\mathrm{H}_{2}} = 4.48 \ \mathrm{eV}$,
$\chi_{\mathrm{H}^{-}} = 0.755 \ \mathrm{eV}$,
and the fitting function of $z_{\mathrm{H}_{2}}$ from \cite{Irwin:1981}.

In Tables~\ref{Planck} and \ref{Rosseland}, we present the numerical values for 
the Planck and Rosseland mean opacities with the equilibrium chemical fractions 
for the temperature and density ranges 
$2.70 < \mathrm{log} \ T \ (\rm K) < 4.30$ and $-17.0 < \mathrm{log}
\ \rho \ (\mathrm{g} \ \mathrm{cm}^{-3}) < -6.0$.  The frequency
integrations in Equations~(\ref{eq:abs-Planck}) and
(\ref{eq:abs-Rosseland}) are carried out in the range of 
$10^{-2} < h\nu/k_{\mathrm{B}} T < 10^{2}$.
\begin{table*}
 \begin{center}
 \caption{Planck mean opacity  log~$\kappa_{\mathrm{P}}$}
 \label{Planck}
  \scalebox{1.05}[1.35]{ 
  \begin{tabular}{|c c c c c c c c c c c c c|} \hline
     & & & & & & $\mathrm{log}~\rho$ & & & & & & \\
    $\mathrm{log}~T$ &-17.0 & -16.0 & -15.0 & -14.0 & -13.0 & -12.0 & -11.0 & -10.0 & -9.0 & -8.0 & -7.0 & -6.0 \\ \hline
    2.7 & -14.928 & -13.898 & -12.898 & -11.898 & -10.898 & -9.898 & -8.898 & -7.898 & -6.898 & -5.898 & -4.898 & -3.898 \\
    2.8 & -14.934 & -13.905 & -12.905 & -11.905 & -10.905 & -9.905 & -8.905 & -7.905 & -6.905 & -5.905 & -4.905 & -3.905 \\
    2.9 & -14.939 & -13.904 & -12.904 & -11.904 & -10.904 & -9.904 & -8.904 &  -7.904 & -6.904 & -5.904 & -4.904 & -3.904 \\
    3.0 & -14.856 & -13.815 & -12.814 & -11.814 & -10.814 & -9.814 &  -8.814 &  -7.814 & -6.814 & -5.814 & -4.814 & -3.814 \\
    3.1 & -15.181 & -13.858 & -12.776 & -11.750 & -10.742 & -9.739 &  -8.738 &  -7.738 & -6.738 & -5.738 & -4.738 & -3.738 \\
    3.2 & -20.250 & -16.670 & -14.552 & -12.483 & -10.944 & -9.758 &  -8.698 &  -7.679 & -6.673 & -5.672 & -4.671 & -3.671 \\
    3.3 & -16.156 & -15.656 & -15.155 & -14.611 & -13.486 & -11.469 & -9.422 &  -7.903 & -6.724 & -5.667 & -4.649 & -3.643 \\
    3.4 & -12.847 & -12.347 & -11.847 & -11.347 & -10.847 & -10.347 & -9.846 &  -9.262 & -7.760 & -6.078 & -4.815 & -3.730 \\
    3.5 & -10.161 & -9.727 &  -9.251 &  -8.759 &  -8.261 &  -7.762 &  -7.263 & -6.763 & -6.266 & -5.793 & -5.292 & -4.094 \\
    3.6 & -6.682 & -6.646 &  -6.576 &  -6.413 &  -6.115 &  -5.705 &  -5.238 &  -4.749 & -4.252 & -3.755 & -3.266 & -2.851 \\
    3.7 & -4.209 & -3.707 &  -3.521 &  -3.456 &  -3.421 &  -3.370 &  -3.248 &  -3.002 & -2.628 & -2.177 & -1.694 & -1.207 \\
    3.8 & -4.495 & -3.494 &  -2.515 &  -1.665 &  -1.182 &  -1.002 &  -0.937 &  -0.896 & -0.825 & -0.664 & -0.368 &  0.039 \\
    3.9 & -4.942 & -3.940 &  -2.940 &  -1.941 &  -0.949 &  -0.023 &   0.617 &   0.888 &  0.983 &  1.033 &  1.104 &  1.258 \\ 
    4.0 & -5.388 & -4.387 &  -3.387 &  -2.387 &  -1.387 &  -0.388 &   0.602 &   1.516 &  2.126 &  2.377 &  2.470 &  2.529 \\ 
    4.1 & -5.835 & -4.833 &  -3.833 &  -2.833 &  -1.833 &  -0.833 &   0.166 &   1.164 &  2.137 &  2.950 &  3.383 &  3.544 \\ 
    4.2 & -6.280 & -5.279 &  -4.279 &  -3.279 &  -2.279 &  -1.279 &  -0.279 &   0.721 &  1.719 &  2.705 &  3.587 &  4.128 \\ 
    4.3 & -6.724 & -5.724 &  -4.724 &  -3.724 &  -2.724 &  -1.724 &  -0.724 &   0.276 &  1.276 &  2.275 &  3.261 &  4.152 \\ \hline
  \end{tabular}
  }
 \end{center}
\end{table*}
\begin{table*}
 \begin{center}
 \caption{Rosseland mean opacity  log~$\kappa_{\mathrm{R}}$}
 \label{Rosseland}
  \scalebox{1.05}[1.35]{ 
  \begin{tabular}{|c | c c c c c c c c c c c c|} \hline
     & & & & & & $\mathrm{log}~\rho$ & & & & & & \\
    $\mathrm{log}~T$ & -17.0 & -16.0 & -15.0 & -14.0 & -13.0 & -12.0 & -11.0 & -10.0 & -9.0 & -8.0 & -7.0 & -6.0 \\ \hline
    2.7 & -11.152 & -11.122 & -10.938 & -10.465 & -9.823 & -9.115 & -8.370 & -7.688 & -7.113 & -6.605 & -6.041 & -5.222 \\ 
    2.8 & -10.753 & -10.740 & -10.638 & -10.259 & -9.636 & -8.925 & -8.189 & -7.531 & -7.002 & -6.513 & -5.839 & -4.918 \\
    2.9 & -10.354 & -10.348 & -10.292 & -10.012 & -9.436 & -8.734 & -8.009 & -7.367 & -6.861 & -6.340 & -5.552 & -4.608 \\
    3.0 & -9.954 & -9.951 &  -9.922 &  -9.733 &  -9.225 & -8.536 & -7.816 & -7.174 & -6.666 & -6.066 & -5.220 & -4.364 \\ 
    3.1 & -9.491 & -9.529 &  -9.529 &  -9.412 &  -8.984 & -8.321 & -7.611 & -6.976 & -6.458 & -5.805 & -5.022 & -4.349 \\ 
    3.2 & -8.764 & -8.863 &  -8.944 &  -9.008 &  -8.798 & -8.142 & -7.410 & -6.761 & -6.216 & -5.555 & -4.917 & -4.394 \\ 
    3.3 & -7.320 & -7.454 &  -7.587 &  -7.715 &  -7.822 & -7.837 & -7.419 & -6.627 & -5.992 & -5.354 & -4.848 & -4.417 \\ 
    3.4 & -5.907 & -6.095 &  -6.267 &  -6.427 &  -6.576 & -6.689 & -6.696 & -6.596 & -6.118 & -5.319 & -4.796 & -4.408 \\ 
    3.5 & -4.143 & -4.495 &  -4.803 &  -5.071 &  -5.305 & -5.506 & -5.635 & -5.613 & -5.491 & -5.329 & -4.812 & -4.325 \\ 
    3.6 & -2.061 & -2.538 &  -2.997 &  -3.425 &  -3.811 & -4.143 & -4.389 & -4.440 & -4.211 & -3.822 & -3.385 & -2.984 \\ 
    3.7 & -0.606 & -0.856 &  -1.260 &  -1.721 &  -2.189 & -2.630 & -2.959 & -2.969 & -2.657 & -2.220 & -1.751 & -1.273 \\
    3.8 & -0.526 & -0.527 &  -0.537 &  -0.611 &  -0.865 & -1.259 & -1.649 & -1.777 & -1.504 & -1.066 & -0.589 & -0.099 \\
    3.9 & -0.526 & -0.526 &  -0.526 &  -0.526 &  -0.529 & -0.556 & -0.686 & -0.832 & -0.653 & -0.227 &  0.253 &  0.745 \\
    4.0 & -0.526 & -0.526 &  -0.526 &  -0.526 &  -0.525 & -0.521 & -0.496 & -0.407 & -0.163 &  0.321 &  0.845 &  1.353 \\
    4.1 & -0.526 & -0.526 &  -0.525 &  -0.524 &  -0.524 & -0.521 & -0.505 & -0.426 & -0.164 &  0.421 &  1.147 &  1.754 \\
    4.2 & -0.526 & -0.526 &  -0.526 &  -0.524 &  -0.518 & -0.513 & -0.504 & -0.455 & -0.265 &  0.228 &  1.064 &  1.911 \\
    4.3 & -0.526 & -0.526 &  -0.526 &  -0.525 &  -0.521 & -0.501 & -0.480 & -0.447 & -0.312 &  0.094 &  0.878 &  1.832 \\ \hline
  \end{tabular}
  }
 \end{center}
\end{table*}

\section{Lithium cooling rate}
\label{App:Li}

We calculate the cooling rate of atomic lithium bound-bound emission 
by counting level populations from the statistical balance among ten (from 2s up to 5s) levels.
Photon trapping effect in large column density cases
is taken into account by way of the escape probability formalism (e.g. \citealt{Omukai:2001}).
The level energies and the spontaneous radiative decay rate are taken from \cite{Klevas:2016}.
We only consider electron impact as the collisional excitation process and its rate coefficients are taken from \cite{Osorio:2011}.
The cooling rate including the line self-absorption 
$\overline{\beta}_{\mathrm{esc}}\Lambda_\mathrm{Li,thin}=n(\mathrm{Li})n(\mathrm{e})L_{\mathrm{Li}}/\rho~\mathrm{(erg ~g^{-1}~s^{-1})}$ 
is calculated for given sets of the temperature $T$, the electron number density $n({\rm e})$ and the lithium column density $N({\rm Li})$. 
We found the results are excellently reproduced by the interpolation in the form of 
\begin{align}
\frac{1}{L_{\mathrm{Li}}} = \frac{1}{L_{0}} + \frac{n(\mathrm{e})}{\overline{\beta}_{\mathrm{esc}}\mathcal{L}_{\mathrm{LTE}}}~,
\label{Eq:Li_fit}
\end{align}
where $L_{0}$ is the cooling function in the low-density limit and 
$\overline{\beta}_{\mathrm{esc}}\mathcal{L}_{\mathrm{LTE}}$ is the cooling rate per lithium atom 
in the LTE case with the photon trapping effect included, 
which are presented in Table~\ref{Tab:LiI_fit} 
as a function of the temperature $T$ and 
the column density parameter $\tilde{N}(\mathrm{Li}) = N(\mathrm{Li})/\sqrt{2k_{\mathrm{B}}T/m_{\mathrm{Li}}}$, 
where $N(\mathrm{Li})$ is the column density of lithium and $m_{ \mathrm{Li}}$ is the mass of a lithium atom. 
The error in this approximation is generally less than one percent and 
at most several percent in the presented range
even in the worst case at high temperatures and the electron number densities.
\begin{table*}
 \begin{center}
 \caption{line cooling parameters for Li I}
 \label{Tab:LiI_fit}
  \scalebox{1.1}[1.2]{ 
  \begin{tabular}{c c c c c c c c c c c} \hline \hline
    \multicolumn{5}{c}{} & \multicolumn{3}{c}{Temperature~~$T$ [K]} & \multicolumn{3}{c}{} \\ \hline
    log~$\tilde{N}$(Li) & & 300 & 500 & 800 & 1000 & 2000 & 3000 & 5000 & 8000 & 10000 \\ \hline
    \multicolumn{4}{c}{} & \multicolumn{5}{c}{log~$L_{0}$ [erg cm$^3$ s$^{-1}$]} & \multicolumn{2}{c}{} \\ \hline
     & & -46.86 & -36.02 & -29.48 & -27.22 & -22.60 & -21.03 & -19.75 & -18.99 & -18.72 \\ \hline
    \multicolumn{4}{c}{} & \multicolumn{5}{c}{log~$\overline{\beta}_{\mathrm{esc}}\mathcal{L}_{\mathrm{LTE}}$ [erg s$^{-1}$]} & \multicolumn{2}{c}{} \\ \hline
    6.0 & & -34.78 & -22.36 & -15.38 & -13.05 & -8.40 & -6.85 & -5.60 & -4.89 & -4.66 \\ 
    7.0 & & -35.64 & -23.23 & -16.25 & -13.92 & -9.26 & -7.71 & -6.44 & -5.64 & -5.36 \\ 
    8.0 & & -36.64 & -24.23 & -17.25 & -14.92 & -10.26 & -8.71 & -7.40 & -6.47 & -6.13 \\ 
    9.0 & & -37.60 & -25.19 & -18.21 & -15.88 & -11.23 & -9.67 & -8.31 & -7.27 & -6.89 \\ 
    10.0 & & -38.28 & -25.87 & -18.88 & -16.56 & -11.90 & -10.35 & -9.05 & -8.13 & -7.78 \\ 
    11.0 & & -38.68 & -26.27 & -19.29 & -16.96 & -12.31 & -10.75 & -9.50 & -8.73 & -8.46 \\ 
    12.0 & & -38.73 & -26.31 & -19.33 & -17.00 & -12.35 & -10.80 & -9.55 & -8.85 & -8.63 \\ 
    13.0 & & -38.73 & -26.32 & -19.34 & -17.01 & -12.35 & -10.80 & -9.56 & -8.87 & -8.65 \\ 
    14.0 & & -38.73 & -26.32 & -19.34 & -17.01 & -12.35 & -10.80 & -9.56 & -8.87 & -8.65 \\ 
    15.0 & & -38.73 & -26.32 & -19.34 & -17.01 & -12.35 & -10.80 & -9.56 & -8.87 & -8.65 \\ \hline
  \end{tabular}
  }
 \end{center}
\end{table*}

\section{Chemical Reactions}
\label{App:inverse}
Our chemical network consists of reactions 1-10 in Table~\ref{reactions}, 
including both forward and reverse reactions. 
The forward (i.e., left to right) reaction coefficients are presented 
in the table. 
The reverse reaction coefficient $k'_{\rm c}$ 
for collisional reactions (reactions 1-8 in Table~\ref{reactions}), 
${\rm A} + {\rm B} \rightleftharpoons {\rm C} +{\rm D}$, 
whose forward reaction coefficient is $k_{\rm c}$, 
is given from the law of mass action as 
\begin{align}
k_{\rm c}' &= k_{\rm c}/K(T)\  ,
\label{eq:chem}
\end{align}
with equilibrium constant $K(T)=\left[ \frac{n({\rm C})n({\rm D})}{n({\rm A}) n({\rm B})} \right]^{*}$
calculated by the Saha-Boltzmann equation. 
The reverse reaction coefficients of radiative reactions (i.e., reactions 9 and 10), 
which are expressed symbolically as ${\rm A + B \rightleftharpoons C + \gamma}$
depend on the radiation field. 
The radiation intensity $J_{\nu}$ in the disc 
is proportional to the effective optical depth for the contiuum $\tau =\sqrt{\tau_{\mathrm{P}}\tau_{\mathrm{R}}}$ 
and saturates to the black-body value $B_{\nu}$ at $\tau \simeq 1$. 
Recalling that the chemical equilibrium is realized for $J_{\nu}=B_{\nu}$, 
the reverse reaction coefficient $k'_{\rm r}$ is related to the forward reaction coefficient $k_{\rm r}$ as  
\begin{align}
k'_{\rm r} = k_{\rm r} \  \mathrm{min} (\tau,1) /K(T),
\label{eq:chem_tau}
\end{align}
where the equilibrium coefficient 
$K(T)=\left[ \frac{n({\rm C})}{n({\rm A}) n({\rm B})} \right]^{*}$. 
\begin{table*}
 \begin{center}
 \caption{Chemical reactions}
 \label{reactions}
  \scalebox{1.0}[1.3]{ 
  \begin{tabular}{c l l c} \hline \hline
    Number & Reaction & Rate coefficient of forward reaction ($\mathrm{cm}^{3} \  \mathrm{s^{-1}}$) & Reference \\ \hline
    1 & $\mathrm{H} + \mathrm{e} \rightleftharpoons \mathrm{H}^{+} + 2 \mathrm{e}$ & $k_{1} = \mathrm{exp}[-3.271396786 \times 10^{1}$ & {\small \cite{Janev:1987}} \\ 
    & & $\  \  \  \  \  \  + 1.35365560 \times 10^{1} \  \mathrm{ln}~T_{\mathrm{e}}$ & \\
    & & $\  \  \  \  \  \  - 5.73932875 \times 10^{0} \  (\mathrm{ln}~T_{\mathrm{e}})^{2}$ & \\
    & & $\  \  \  \  \  \  + 1.56315498 \times 10^{0} \  (\mathrm{ln}~T_{\mathrm{e}})^{3}$ & \\
    & & $\  \  \  \  \  \  - 2.87705600 \times 10^{-1} \  (\mathrm{ln}~T_{\mathrm{e}})^{4}$ & \\
    & & $\  \  \  \  \  \  + 3.48255977 \times 10^{-2} \ (\mathrm{ln}~T_{\mathrm{e}})^{5}$ & \\
    & & $\  \  \  \  \  \  - 2.63197617 \times 10^{-3} \  (\mathrm{ln}~T_{\mathrm{e}})^{6}$ & \\
    & & $\  \  \  \  \  \  + 1.11954395 \times 10^{-4} \ (\mathrm{ln}~T_{\mathrm{e}})^{7}$ & \\
    & & $\  \  \  \  \  \  - 2.03914985 \times 10^{-6} \  (\mathrm{ln}~T_{\mathrm{e}})^{8}]$ & \\
    2 & $\mathrm{H}^{-} + \mathrm{H} \rightleftharpoons \mathrm{H}_{2} + \mathrm{e}$ & $k_{2} = 1.3500\times10^{-9}( T^{9.8493\times10^{-2}}+3.2852\times10^{-1}T^{5.5610\times10^{-1}}$ & {\small \cite{Kreckel:2010}} \\    
    & & $\  \  \  \  \  \  +2.7710\times10^{-7}T^{2.1826})/(1.0+6.1910\times10^{-3}T^{1.0461}$ & \\
    & & $\  \  \  \  \  \  +8.9712\times10^{-11}T^{3.0424})+3.2576\times10^{-14}T^{3.7741})$ & \\    
    3 & $2 \mathrm{H} \rightleftharpoons \mathrm{H}^{+} + \mathrm{e} + \mathrm{H}$ & $k_{3} = 1.2\times10^{-17}T^{1.2}~\mathrm{exp}\left( -\frac{157800}{T} \right)$ & {\small \cite{Lenzuni:1991}} \\         
    4 & $\mathrm{H}_{2} + \mathrm{e} \rightleftharpoons 2 \mathrm{H} + \mathrm{e}$ & $k_{4} = k_{4, \mathrm{H}}^{1-a} k_{4, \mathrm{L}}^{a} $ & \\
    & & $k_{4, \mathrm{H}} = 1.91 \times 10^{-9} T^{0.136} \mathrm{exp}\Bigl( - 53407.1/T \Bigr)$ & {\small \cite{Trevisan:2002}} \\
    & & $k_{4, \mathrm{L}} = 4.49 \times 10^{-9} T^{0.11} \mathrm{exp}\Bigl( - 101858/T \Bigr)$ & {\small \cite{Trevisan:2002}} \\ 
    & & $a = \left( 1 + n_{\mathrm{H}} / n_{\mathrm{crit}} \right)^{-1}$ & \\        
    & & $n_{\mathrm{crit}} = y(\mathrm{H})/n_{\mathrm{crit}}(\mathrm{H}) + 2y(\mathrm{H}_{2})/n_{\mathrm{crit}}(\mathrm{H}_{2}) + y(\mathrm{He})/n_{\mathrm{crit}}(\mathrm{He})$ & \\            
    & & $\mathrm{log}~(n_{\mathrm{crit}}(\mathrm{H})) = 3 - 0.416 \  \mathrm{log}~(T/10^{4}) - 0.372 \left[ \mathrm{log}~(T/10^{4}) \right]^{2}$ & \\                
    & & $\mathrm{log}~(n_{\mathrm{crit}}(\mathrm{H}_{2})) = 4.845 - 1.3 \  \mathrm{log}~(T/10^{4}) + 1.62 \left[ \mathrm{log}~(T/10^{4}) \right]^{2}$ & \\                
    & & $\mathrm{log}~(n_{\mathrm{crit}}(\mathrm{He})) = 5.0792 \left[ 1 - 1.23 \times 10^{-5} (T - 2000) \right]$ & \\            
    5 & $\mathrm{H}_{2} + \mathrm{H} \rightleftharpoons 3 \mathrm{H}$ & see the reference & {\small \cite{Martin:1996}} \\
    6 & $\mathrm{H}^{-} + \mathrm{H}^{+} \rightleftharpoons 2 \mathrm{H}$ & $k_{6} = 2.4 \times 10^{-6} T^{-0.5} \Bigl( 1.0 + T/20000 \Bigr)$ & {\small \cite{Croft:1999}} \\
    7 & $2 \mathrm{H}_{2} \rightleftharpoons 2 \mathrm{H} + \mathrm{H}_{2}$ & $k_{7} = k_{7, \mathrm{H}}^{1-a} k_{7, \mathrm{L}}^{a} $ & \\
    & & $k_{7, \mathrm{H}} = 1.3 \times 10^{-9} \mathrm{exp} \Bigl( - 53300/T \Bigr)$ & {\small \cite{Shapiro:1987}} \\
    & & $k_{7, \mathrm{L}} = \frac{5.996 \times 10^{-30} T^{4.1881}}{(1.0+6.761\times 10^{-6} T)^{5.6881}} \mathrm{exp} \Bigl( - 54657.4/T \Bigr)$ & {\small \cite{Martin:1998}} \\
    8 & $\mathrm{H}_{2} + \mathrm{He} \rightleftharpoons 2 \mathrm{H} + \mathrm{He}$ & $k_{8} = k_{8, \mathrm{H}}^{1-a} k_{8, \mathrm{L}}^{a} $ & \\
    & & $k_{8, \mathrm{H}} =  \mathrm{dex}[-1.75~\mathrm{log}T -2.729 - 23474/T]$ & {\small \cite{Dove:1987}} \\
    & & $k_{8, \mathrm{L}} = \mathrm{dex}[3.801~\mathrm{log}T -27.029 - 29487/T]$ & {\small \cite{Dove:1987}} \\
    9 & $\mathrm{H}^{+} + \mathrm{e} \rightleftharpoons \mathrm{H} + \gamma$ & $k_{9} = 2.753 \times 10^{-14} \Bigl( 315614/T \Bigr)^{1.5}$ & {\small \cite{Ferland:1992}} \\
    & & $\  \  \  \  \  \  \times \Bigl[ 1.0 + \Bigl( 115188/T \Bigr)^{0.407} \Bigr]^{-2.242}$ & \\
    10 & $\mathrm{H} + \mathrm{e} \rightleftharpoons \mathrm{H}^{-} + \gamma$ & $k_{10} = \mathrm{dex}[-17.845 + 0.762 \mathrm{log}~T$ & {\small \cite{Wishart:1979}} \\
    & & $\  \  \  \  \  \  + 0.1523 (\mathrm{log}~T)^{2}$ & \\
    & & $\  \  \  \  \  \  - 0.03274 (\mathrm{log}~T)^{3}]$ \  \  \  \  \  \  \  \  ($T < 6000 \  \mathrm{K}$)& \\
    & & $\  \  \  = \mathrm{dex}[-16.4199 + 0.1998 (\mathrm{log}~T)^{2}$ & \\
    & & $\  \  \  \  \  \  - 5.447 \times 10^{-3} (\mathrm{log}~T)^{4}$ & \\
    & & $\  \  \  \  \  \  + 4.0415 \times 10^{-5} (\mathrm{log}~T)^{6}]$ \  ($T > 6000 \  \mathrm{K}$)& \\ \hline
    \multicolumn{2}{l}{{\footnotesize Note.--- The temperature $T_{\mathrm{e}}$ is in eV.}} & \multicolumn{1}{c}{} \\
  \end{tabular}
  }
  \\
 \end{center}
\end{table*}


\bsp	
\label{lastpage}
\end{document}